\documentclass[12pt]{article}

\usepackage{subfigure}
\usepackage{amssymb}
\usepackage{amsfonts}
\usepackage{bm}
\usepackage{color}
\usepackage{calc}
\usepackage[latin1]{inputenc}
\usepackage{amsfonts}
\usepackage{amsmath}
\usepackage{mathtools}
\usepackage[english]{babel}
\usepackage[T1]{fontenc}
\usepackage{verbatim}
\usepackage{subfig}
\usepackage{enumerate}
\usepackage{graphicx}
\usepackage{natbib}
\usepackage{url}
\usepackage{algorithmic,algorithm}
\usepackage{placeins}

\def\Bmp#1{ \begin{minipage}{#1} }
\def\Bmpc#1{ \begin{minipage}[c]{#1} }
\def\Bmpt#1{ \begin{minipage}[t]{#1} }
\def\Bmpb#1{ \begin{minipage}[b]{#1} }
\def\Emp{ \end{minipage} }
\newcommand{\revt}[1]{{\color{black}#1}}

\newcommand{\A}{\bm{A}}
\newcommand{\B}{\bm{B}}
\newcommand{\C}{\bm{C}}
\newcommand{\bF}{\bm{F}}
\newcommand{\G}{\bm{G}}
\newcommand{\K}{\bm{K}}
\newcommand{\bP}{\bm{P}}

\newcommand{\Q}{\bm{Q}}

\newcommand{\X}{\bm{X}}
\newcommand{\F}{\mathcal{F}}

\def\J{{\mathcal{J}}}

\def\O{\mbox{\textit{O}}}
\def\R{{\mathcal{R}}}

\def\Xn{{\bm X}_n}

\def\CC{{\mathbb{C}}}

\def\RR{{\mathbb{R}}}

\def\ZZ{{\mathbb{Z}}}

\def\bH{{\bm H}}

\def\b{{\bf b}}

\def\g{{\bf g}}
\def\h{{\bf h}}

\def\0{{\bf 0}}

\def\tb{\widetilde{{\bf b}}}

\def\hg{\widehat{{\bf g}}}

\newcommand{\tB}{\widetilde{\bm{B}}}

\def\bsigma{\boldsymbol{\sigma}}

\def\bxi{\boldsymbol{\xi}}
\def\bbeta{\boldsymbol{\eta}}

\def\hzeta{{\widehat{\zeta}}}

\def\Dpartial#1#2{ {\frac{\partial #1}{\partial #2} }}

\newcommand{\diag}{\operatorname{diag}}

\begin{document}
\title{Harnessing the Kelvin-Helmholtz Instability: Feedback Stabilization of an Inviscid Vortex Sheet}

\author{Bartosz Protas$^{1, }$\thanks{Email address for correspondence: bprotas@mcmaster.ca} \ and Takashi Sakajo$^{2}$
\\ \\ 
$^1$ Department of Mathematics and Statistics, McMaster University \\
Hamilton, Ontario, L8S 4K1, Canada
\\ \\
$^2$ Department of Mathematics, Kyoto University \\ 
Kitashirakawa Oiwake-cho, Sakyo-ku, Kyoto, 606-8502, Japan
}

\date{\today}

\maketitle

\begin{abstract}
   In this investigation we use a {simple model of the dynamics of an
    inviscid vortex sheet} given by the Birkhoff-Rott equation to
  {obtain} fundamental insights about the potential for stabilization
  of shear layers using feedback \revt{control}.  \revt{As actuation
    we consider two arrays of point sinks/sources located a certain
    distance above and below the vortex sheet and subject to the
    constraint that their mass fluxes separately add up to zero.}
  First, we demonstrate using analytical computations that the
  Birkhoff-Rott equation linearized around the flat-sheet
  configuration is in fact controllable when \revt{the number of
    actuator pairs is sufficiently large relative to the number of
    discrete degrees of freedom present in the system, a result valid
    for generic actuator locations.}  Next we design a state-based LQR
  stabilization strategy where the key difficulty is the numerical
  solution of the Riccati equation in the presence of severe
  ill-conditioning resulting from the properties of the Birkhoff-Rott
  equation and the chosen form of actuation, an issue which is
  overcome by performing computations with {a} suitably increased
  arithmetic precision. Analysis of the linear closed-loop system
  reveals exponential decay of the perturbation energy \revt{and of
    the corresponding \revt{actuation energy} in all cases.}
  Computations performed for the nonlinear closed-loop system
  demonstrate that initial perturbations of nonnegligible amplitude
  can be effectively stabilized when a sufficient number of actuators
  is used. We also thoroughly analyze the sensitivity of the
  closed-loop stabilization strategies to the variation of a number of
  key parameters. Subject to the known limitations of inviscid vortex
  models, our findings indicate that, in principle, it may be possible
  to stabilize shear layers \revt{for relatively large initial
    perturbations, provided the actuation has sufficiently many
    degrees of freedom.}
\end{abstract}

\begin{flushleft}
Keywords:
Vortex instability, Instability control, Shear layers
\end{flushleft}


\section{Introduction}
\label{sec:intro}

Shear layers play a prominent role in fluid mechanics as canonical
models of separated flows. Under typical conditions, they undergo the
Kelvin-Helmholtz instability during which the vorticity in the shear
layer becomes concentrated in large vortex structures. Because of this
universal feature, shear layers often serve as models for more
complicated vortex-dominated flows arising in a broad range of
applications.  Already for a long time their properties have
been investigated at the fundamental level from the experimental
\citep{m11a}, theoretical \citep{m86a} as well as the numerical
standpoint \citep{cll92}. In addition to offering insights about
transition to turbulence and the role played in this multiscale
process by vortex structures, shear layers have also served as a
playground for the application of various flow-control techniques
aiming to stabilize the flow by preventing the growth of vortex
structures. While the literature on this topic is quite rich, as
examples in this context, we mention the recent investigations by
\citet{Pastoor2008jfm,Kaiser_etal_14} and by \citet{Wei2009jfm} which
combine experimental, theoretical and numerical approaches. In the
present study, we address the problem of stabilizing a shear layer at
the fundamental level by combining a simple inviscid model of this
phenomenon with methods of modern linear control theory.

Inviscid vortex models using singular vorticity distributions to
represent flow fields have had a long and rich history in fluid
mechanics. Our attention here will be focused exclusively on
two-dimensional (2D) problems. The most common models of this type are
point vortices where the vorticity has the form of a collection of
Dirac delta distributions with time-evolving locations governed by a
system of nonlinear ODEs \citep{NewtonBook}. While point vortices are
typically (albeit not exclusively) employed to model compact vorticity
distributions, continuous vorticity distributions, such as those
arising in shear layers, are represented more accurately in terms of
``vortex sheets'' where the singular vorticity distribution has the
form of a continuous one-dimensional (1D) open or closed curve. In
such models the linear circulation density plays the role of
``vorticity'' (which is not defined on the sheet) and can be
interpreted in terms of the jump of the tangential velocity component
across the sheet. The evolution of the sheet is governed by the
Birkhoff-Rott equation which is a singular integro-differential
equation \citep{fundam:saffman1}.  Vortex sheets are subject to the
Kelvin-Helmholtz instability: infinitesimal perturbations with
wavenumber $n$ superimposed on a flat vortex sheet grow exponentially
in time at a rate proportional to $\vert n \vert$. Not only does this
high-wavenumber instability give rise to ill-posedness in the sense of
Hadamard, but, reinforced through nonlinear effects, it also causes
the development of a finite-time curvature singularity in the sheet
before large localized vortex structures can
appear~\citep{vsheet:Moore}. As a result, computational studies
involving the Birkhoff-Rott equation typically require some
regularization in order to track its long-time evolution, usually in
the form of the well-known ``vortex-blob'' approach \citep{k86a} or
using the more recent Euler-alpha strategy \citep{hnp06a}.  A generic
feature characterizing the evolution of (regularized) inviscid vortex
sheets is roll-up producing localized vortex spirals \citep{k86b}. It
should be pointed out that the appearance of such rolling-up vortex
spirals is closely related to the emergence of the curvature
singularity {in the solutions of the regularized equation with
  complexified time} \citep{vsheet:Sa04}.  In addition, interaction of
these vortex spirals leads to chaotic evolution \citep{vsheet:KrNi02}
and complex mixing \citep{vsheet:SaOk98} marking the onset of
turbulence in the shear layers.  To summarize, it is fair to say that
the Kelvin-Helmholtz instability acts as a key trigger for the
emergence of turbulent mixing shear layers. Let us also note that
there are many interesting mathematical questions concerning various
properties of solutions to the Birkhoff-Rott equation and we refer the
reader to the collection edited by \citet{c89a} and the monograph by
\citet{mb02} for further details on this topic.  In the present study
we will focus on a flat periodic vortex sheet as an idealized model of
a shear layer and will investigate how various forms of actuation
combined with methods of the modern linear control theory can be used
to overcome the Kelvin-Helmholtz instability and thus inhibit
emergence of turbulent shear layers.  We will consider here an
arguably more difficult case when there is no regularization in the
Birkhoff-Rott equation.

The potential offered by the mathematical methods of control theory to
solve various flow-control problems was recognized already in the
1990s. Two main classes of problems are of interest: open-loop, or
off-line, control problems where one seeks to optimize certain inputs
such as, e.g., actuation through a particular form of boundary
conditions, and closed-loop control problems in which one typically
attempts to stabilize the problem against disturbances through some
form of state-based or output-based feedback. Problems in the first
category usually have the form of PDE-constrained optimization and are
often solved using variants of the adjoint-based method. On the other
hand, problems in the second category require determination of
suitable feedback operators {(kernels)} which in the linear
setting is accomplished by solving some form of the operator Riccati
equation.  Since the seminal works of \citet{bmt01} and
\citet{Bewley1998jfm}, both approaches have seen remarkable success in
solving a range of diverse flow-control problems of both fundamental
and applied significance. While reviewing the corresponding literature
is beyond the scope of this study, we refer the reader to the
monograph by \citet{g03} and to the survey papers by
\citet{Kim2007arfm,bhhs09,bn15a} for details. Another class of
problems which rely on similar mathematical techniques is flow
estimation based on incomplete and possibly noisy measurements. We
remark that, in order to produce approaches applicable in real-life
situations, flow-control techniques usually have to be combined with
state estimation. Finally, due to their special mathematical
structure, various vortex models offer particular opportunities for
flow control and progress along these lines was reviewed by
\citet{p08a}.

The goal of the present study is first to provide a precise
control-theoretic characterization of the inviscid vortex sheet model
for various types of actuation. In this context we {demonstrate} that
the linearization of the Birkhoff-Rott equation around the flat-sheet
configuration is in fact controllable when the actuation
\revt{involves a sufficient number of point sinks/sources or point
  vortices located} on each side of the sheet and at a certain
distance from it. Then, we design a state-feedback stabilization
strategy by synthesizing a linear-quadratic regulator (LQR) and assess
its performance as a function of a number of different parameters in
both the linear and nonlinear regime. Due to certain peculiar
properties of the Birkhoff-Rott equation and the considered form of
actuation, computational solution of this problem requires a special
approach relying on calculations performed with a very high arithmetic
precision. Rather than design a control strategy applicable in actual
experiments, our study aims to provide fundamental insights about
opportunities and limitations inherent in the feedback control of the
Kelvin-Helmholtz instability. The structure of the paper is as
follows: in the next section we introduce the Birkhoff-Rott equation
as a model of a shear layer and discuss its linearization as well as
the form of the control actuation; in \S \ref{sec:control} we present
a control-theoretic characterization of this model and introduce the
LQR stabilization strategy; details concerning our numerical approach
are discussed in \S \ref{sec:numer}, whereas computational results are
presented in \S \ref{sec:results}; discussion of our findings and
final conclusions are deferred to \S \ref{sec:final}. Some additional
technical material is collected in two appendices.

\section{Inviscid Vortex Sheets}
\label{sec:sheet}

\begin{figure}
\centering
\includegraphics[width=0.7\textwidth]{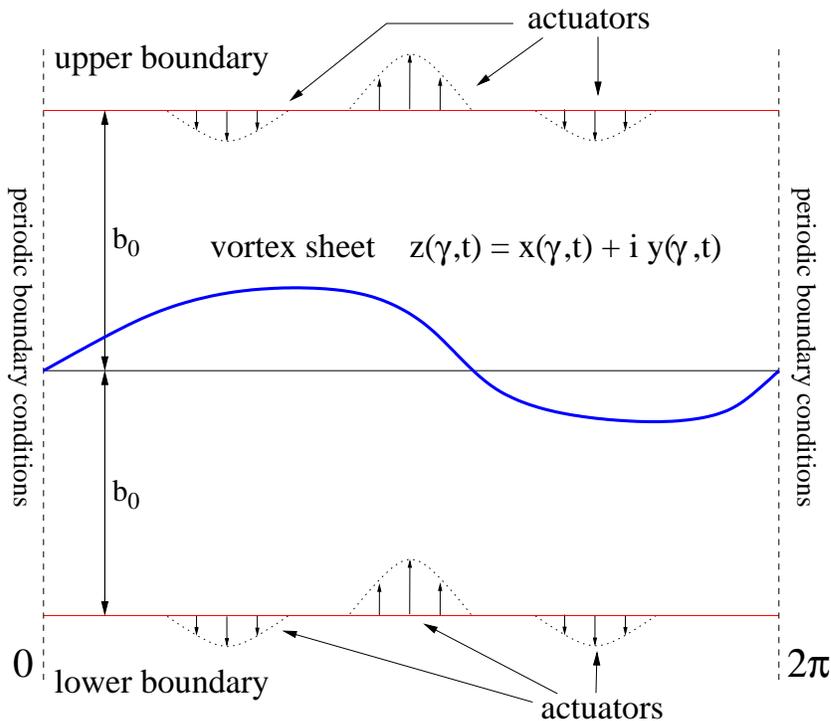}
\caption{\revt{Schematic illustration of the problem set-up. The
    domain is $2\pi$-periodic in the horizontal direction and the
    vortex sheet at time $t$ is represented by
    $z(\gamma,t)=x(\gamma,t)+ i y(\gamma,t)$, in which $\gamma$ is a
    parameter varying along the sheet. An array of actuators is placed
    on a {\em hypothetical} boundary above and below the sheet.}}
\label{fig:setting}
\end{figure}

In this section we introduce a mathematical model for an inviscid
vortex sheet in an unbounded domain which, for simplicity, will be
assumed to be periodic in the horizontal direction with the period of
$2\pi$.  \revt{In \S \ref{sec:actuation} below we discuss how this
  model can be modified to account for the presence of actuators.}  We
assume that the sheet has a constant intensity equal to one. As is
common in the study of such problems \citep{fundam:saffman1}, we will
extensively use the complex representation of different quantities and
will identify a point $(x,y)$ in the 2D space with $z = x + iy$ in the
complex plane, where $i$ is the imaginary unit.  Let then
$z(\gamma,t)$ denote the position of a point on the sheet which
corresponds to the circulation parameter $\gamma \in [0,2\pi]$ and
some time $t$. The quantity $\gamma$ {represents} a way of
parameterizing the sheet and for sheets of constant intensity is
proportional to the arc-length of the curve. Periodicity of the sheet
then implies
\begin{equation}
z(\gamma+ 2 \pi, t)=z(\gamma,t)+2\pi, \quad \gamma \in [0,2\pi].
\label{eq:z}
\end{equation}
\revt{The flow set-up is shown schematically in figure
  \ref{fig:setting}.}  Under the assumptions of periodicity and
constant unit strength, the Birkhoff-Rott equation describing the
evolution of the vortex sheet takes the form
\begin{equation}
\Dpartial{z^\ast}{t}(\gamma,t) = V(z(\gamma,t))
\revt{\,\equiv\,} \frac{1}{4\pi i}\mbox{pv} \int_0^{2\pi} 
\cot\left( \frac{z(\gamma,t)-z(\gamma^\prime,t) }{2} \right)\,d\gamma^\prime, 
\label{eq:BR}
\end{equation}
where $z^\ast$ denotes the complex conjugate of $z$, the integral on
the right-hand side (RHS) is understood in Cauchy's principal-value
sense and $V(z) = (u-iv)(z)$ represents the complex velocity at the
point $z$ with $u$ and $v$ the horizontal and vertical velocity
components, respectively. Equation \eqref{eq:BR} is complemented with a
suitable initial condition $z(\gamma,0) = z_0(\gamma)$, $\gamma \in
[0,2\pi]$. As can be readily verified, equation \eqref{eq:BR} admits
an equilibrium solution (a fixed point) $\tilde{z}(\gamma,t) = \gamma$
for $\gamma \in [0,2\pi]$, i.e.,
$\Dpartial{}{t}\tilde{z}^\ast(\gamma,t) = 0$, which corresponds to a
flat (undeformed) sheet. The linear stability of this equilibrium
configuration will be discussed below.

\subsection{Kelvin-Helmholtz Instability}
\label{sec:KH}

\begin{figure}
\centering
\mbox{
\subfigure[]{\includegraphics[width=0.5\textwidth]{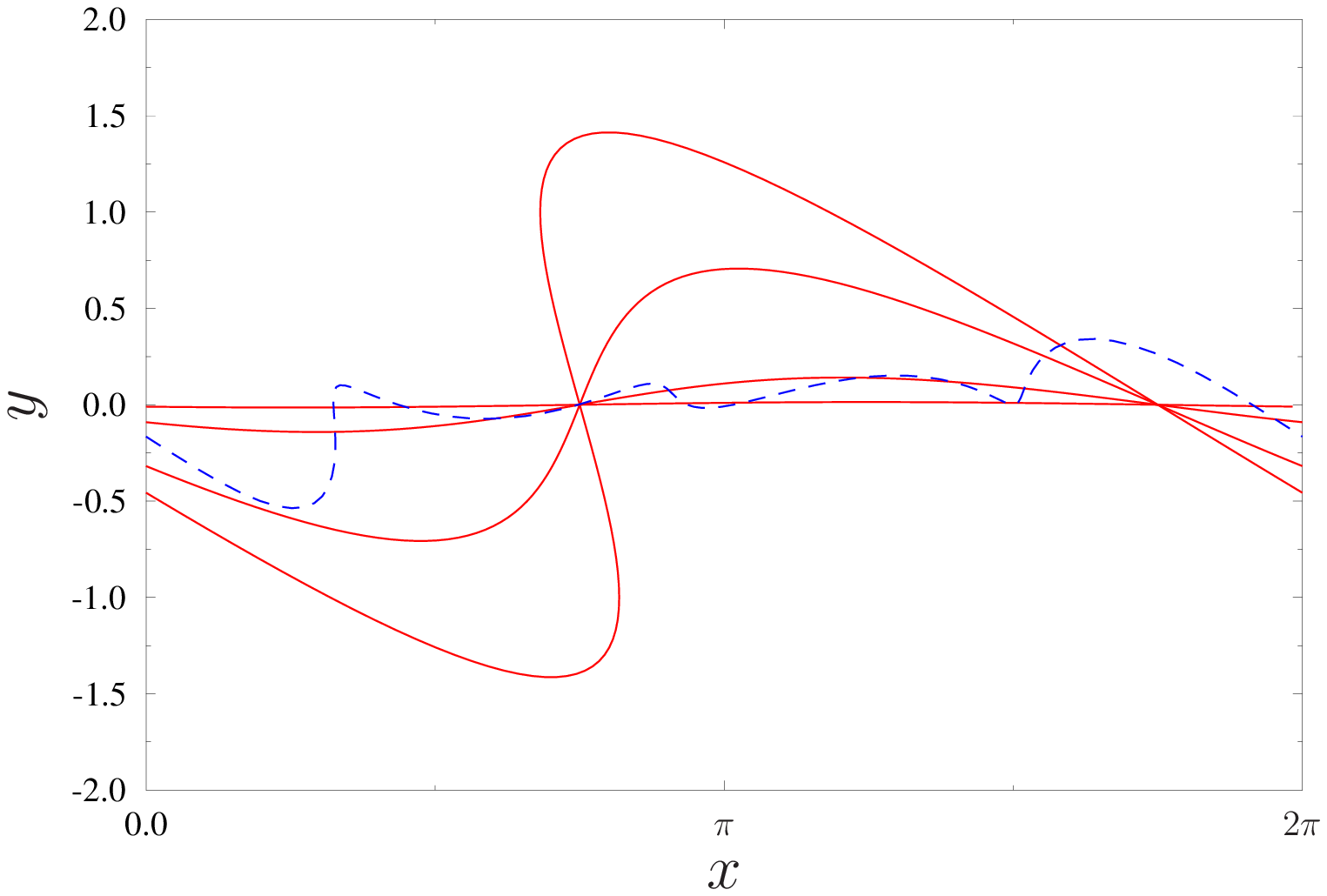}}
\qquad
\subfigure[]{\includegraphics[width=0.5\textwidth]{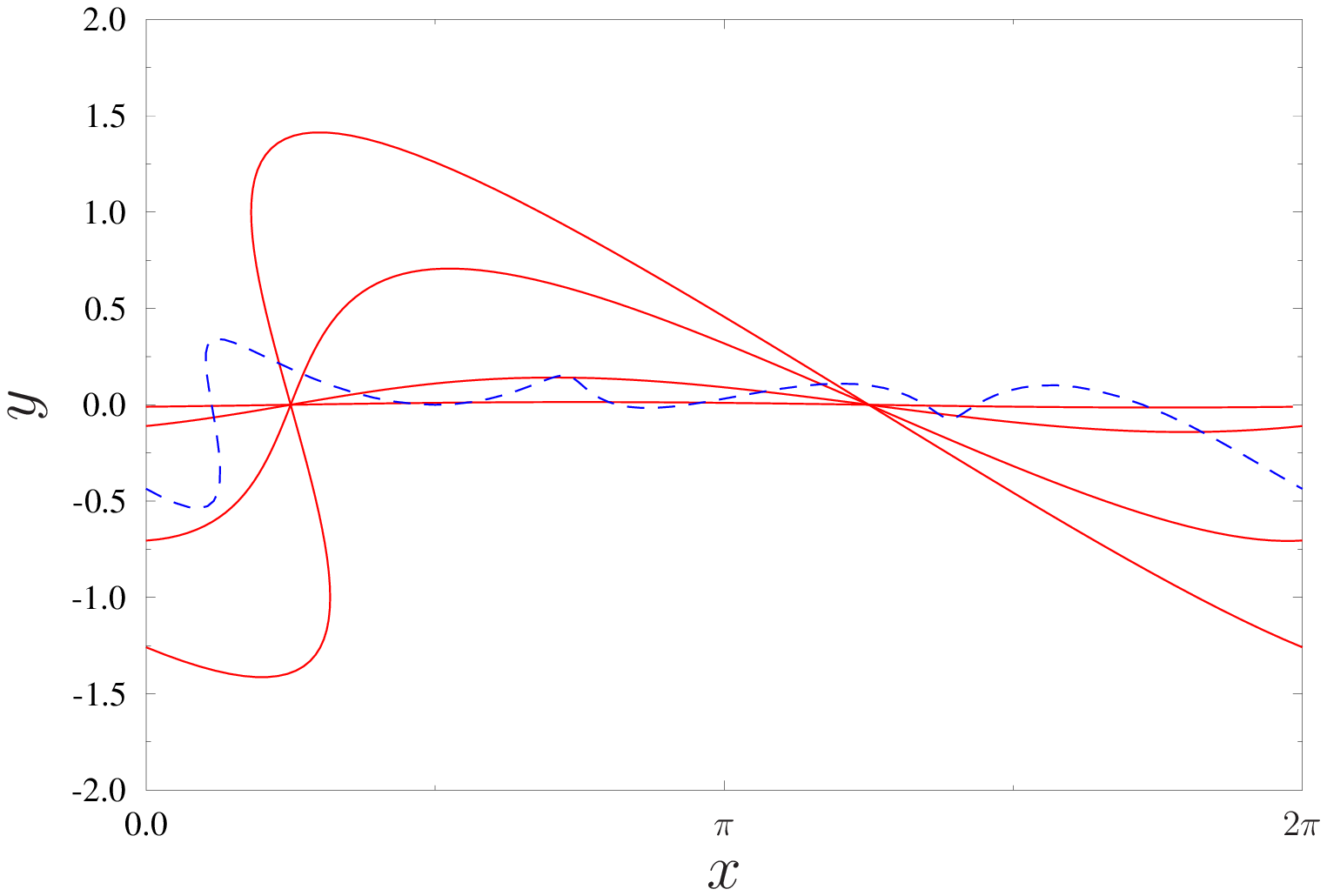}}}
\caption{Equilibrium configuration $\tilde{z}$ perturbed with the
  unstable eigenvectors \revt{(a) $\xi_1$ and (b) $\xi_2$ for (red
    solid lines) $n=1$ and $\varepsilon = 0.01$, $0.1$, $0.5$ and
    $1.0$, cf.~\eqref{eq:zpert}; blue dashed lines represent
    equilibrium configurations with perturbations obtained by
    superposing components $\xi_1$ and $\xi_2$ with different
    wavenumbers $n=1,2,3,4$ and the same value $\varepsilon = 0.1$.}}
\label{fig:xi}
\end{figure}

We now characterize the Kelvin-Helmholtz instability undergone by the
flat-sheet equilibrium configuration. This will be done by performing
a linear-stability analysis of the Birkhoff-Rott equation in the
neighbourhood of the fixed point $\tilde{z}$.  Following the approach
of \citet{vsheet:SaOk96}, see also \citet{ka79a}, suppose that we
perturb this equilibrium state infinitesimally as
\begin{equation}
z(\gamma, t)=\gamma + \varepsilon \zeta(\gamma,t),
\label{eq:zpert}
\end{equation}
for $0 < \varepsilon \ll 1$. Here, $\zeta(\gamma,t)$ is a perturbation
represented in the periodic setting, cf.~\eqref{eq:z}, as
\begin{equation}
\zeta(\gamma,t) =  \sum_{n=-\infty}^\infty \hzeta_n(t) \mbox{e}^{i n \gamma},
\label{eq:zeta}
\end{equation}
in which $\hzeta_n \in \CC$, $n \in \ZZ$, are the Fourier
coefficients.  Then, we obtain the linearized equation for
$\zeta(\gamma,t)$ as follows.
\begin{eqnarray}
\frac{\partial \zeta^\ast}{\partial t}(\gamma,t) &=& -\frac{1}{8 \pi i} \mbox{pv} \int_0^{2\pi} \frac{\zeta(\gamma,t)-\zeta(\gamma^\prime,t)}{\sin^2\left( \frac{\gamma-\gamma^\prime}{2} \right)} \,d\gamma^\prime  \nonumber \\
 &=& \sum_{n=-\infty}^\infty \hzeta_n(t) \mbox{e}^{i n \gamma} \left[ -\frac{1}{8\pi i} \mbox{pv} \int_0^{2\pi} \frac{1-\mbox{e}^{-in \gamma^\prime}}{\sin^2\left(\frac{\gamma^\prime}{2}\right)}\, d\gamma^\prime  \right] \nonumber \\
 &=& \sum_{n=-\infty}^\infty \hzeta_n(t) \mbox{e}^{i n \gamma} \left[ -\frac{1}{4 i} \mbox{pv} \int_0^1 \frac{1-\cos(2 \pi n \tau)-i \sin(2\pi n \tau)}{\sin^2(\pi \tau)}\, d\tau  \right] \nonumber \\
 &=& -\frac{1}{2i} \sum_{n=1}^\infty n \, \hzeta_n(t) \mbox{e}^{i n \gamma} -\frac{1}{2i} \sum_{n=1}^\infty  n \,\hzeta_{-n}(t) \mbox{e}^{-i n \gamma}. \label{eq:BRlin}
\end{eqnarray}
Since
\begin{equation}
\frac{\partial \zeta^\ast}{\partial t} = \sum_{n=-\infty}^\infty \frac{\mbox{d} \hzeta^\ast_n}{\mbox{d}t} \mbox{e}^{-in\gamma}
 = \sum_{n=1}^\infty \frac{\mbox{d} \hzeta_n^\ast}{\mbox{d}t} \mbox{e}^{-in\gamma} + \sum_{n=1}^\infty \frac{\mbox{d} \hzeta_{-n}^\ast}{\mbox{d}t} \mbox{e}^{in\gamma} + \frac{\mbox{d}\hzeta_0}{\mbox{d} t},
\label{eq:dzetadt0}
\end{equation}
equating coefficients of the Fourier components in \eqref{eq:BRlin}
and \eqref{eq:dzetadt0} corresponding to different wavenumbers $n$, we
obtain linearized equations for the coefficients $\hzeta_n$ in a
block-diagonal form
\begin{subequations}
\label{Leq}
\begin{align}
\frac{\mbox{d} \hzeta_{-n}^\ast}{\mbox{d}t} &= \phantom{-} \frac{i n}{2} \hzeta_n,  \label{Leq1}\\
\frac{\mbox{d} \hzeta_n}{\mbox{d}t} &= -\frac{i n}{2} \hzeta_{-n}^\ast, \label{Leq2} \\
\frac{\mbox{d} \hzeta_0}{\mbox{d} t} &= \phantom{-} 0, \label{Leq3}
\end{align}
\end{subequations}
where $n \ge 1$. This step thus allows us to convert the original
continuous PDE problem to an infinite system of ODEs. Since this
latter system has a block-diagonal structure, key insights about the
infinite-dimensional problem can be obtained by analyzing just
  a single block, independently from all other blocks.  The
eigenvalues corresponding to each diagonal block with $n \ge 1$,
cf.~\eqref{Leq1}--\eqref{Leq2}, are $\lambda_n = \pm \frac{n}{2}$.
Since the system must be recast in a purely real form for our
subsequent analysis of its controllability in \S \ref{sec:control}, we
need to rewrite the linearized equations \eqref{Leq1} and \eqref{Leq2}
in terms of the real and imaginary parts of the Fourier coefficients
$\hzeta_n= \alpha_n + i\beta_n$ and $\hzeta_{-n} = \alpha_{-n}+ i
\beta_{-n}$ as
\begin{subequations}
\label{eq:dabdt}
\begin{alignat}{2}
\frac{\mbox{d}}{\mbox{d}t}(\alpha_{-n} - i \beta_{-n}) &=  \phantom{=} \frac{i n}{2} (\alpha_n+i\beta_n) & 
 & = -iB_{-n}(\alpha_n+i \beta_n), \label{eq:dabdta} \\
\frac{\mbox{d}}{\mbox{d}t}(\alpha_n + i \beta_n) &=  -\frac{i n}{2} (\alpha_{-n}-i\beta_{-n})&&  = \phantom{-} i B_n (\alpha_{-n} - i \beta_{-n}), \label{eq:dabdtb}
\end{alignat}
\end{subequations}
where 
\begin{equation*}
B_{-n} = B_n = -\frac{n}{2}, \quad n \ge 1.
\end{equation*}
Hereafter, the real and imaginary parts of the Fourier coefficients
will serve as our state variables. Defining the vector of the state
variables associated with the wavenumber $n$ as $\Xn = [ \alpha_{-n} \
\beta_n \ \beta_{-n} \ \alpha_n ]^T$, relations \eqref{eq:dabdt} can
be recast in the following form
\begin{equation}
\frac{\mbox{d}}{\mbox{d}t}\left[ \begin{array}{c} \alpha_{-n} \\ \beta_n \\ \beta_{-n} \\ \alpha_n  \end{array} \right] = 
-\frac{{n}}{2}
\left[ \begin{array}{cccc} 0 & 1 & 0 & 0 \\  1 & 0 & 0 & 0 \\ 
0 & 0 & 0 & 1 \\ 0 & 0 & 1 & 0 \end{array}\right]
\left[ \begin{array}{c} \alpha_{-n} \\ \beta_n \\ \beta_{-n} \\ \alpha_n  \end{array} \right],
\label{eq:A}
\end{equation}
\revt{which} can be rewritten more compactly as
\begin{equation}
 \frac{\mbox{d}{\bm X}_n}{\mbox{d}t} =  -\frac{n}{2}
{\A_0} {\bm X}_n  = {\bm A}_n {\bm X}_n, \quad n \ge 1,
\label{eq:A2}
\end{equation}
where  $\A_n = -\frac{n}{2} {\A_0}$ and
\begin{equation*}
{\A_0} = \begin{bmatrix}
\C & 0 \\ 0 & \C 
\end{bmatrix}
\quad \text{with} \quad
\C = \begin{bmatrix}
0 & 1 \\ 1 & 0
\end{bmatrix}.
\end{equation*}
It is well known that the stability properties of linear systems with
block-diagonal structure are determined by the eigenvalues and
eigenvectors of the individual blocks.  We thus observe that, for each
wavenumber $n \ge 1$, the evolution described by system \eqref{eq:A2}
is characterized by two invariant subspaces spanned by \revt{the
  vectors $[\alpha_{-n} \ \beta_n \ 0 \ 0 ]^T$ and $[ 0 \ 0 \
  \beta_{-n} \ \alpha_n ]^T$}, which are orthogonal to each other.
Each of these subspaces contains a growing and a decaying mode with
the growth rates given by the eigenvalues $\lambda_n^+ = n / 2$ and
$\lambda_n^- = - n / 2$.  Therefore, for each wavenumber $n \ge 1$, we
have two unstable modes proportional to
\begin{subequations}
\label{eq:hxi12}
\begin{align}
\widehat{\bxi}_1 & = [ -1 \ 1 \ 0 \ 0 ]^T, \label{eq:hxi1} \\ 
\widehat{\bxi}_2 & = [ 0 \ 0 \ -1 \ 1 ]^T, \label{eq:hxi2} 
\end{align}
\end{subequations}
and two stable modes proportional to 
\begin{subequations}
\label{eq:hxi34}
\begin{align}
\widehat{\bxi}_3 & = [ 1 \ 1 \ 0 \ 0 ]^T, \label{eq:hxi3} \\ 
\widehat{\bxi}_4 & = [ 0 \ 0 \ 1 \ 1 ]^T, \label{eq:hxi4} 
\end{align}
\end{subequations}
where $\widehat{\bxi}_1, \widehat{\bxi}_2, \widehat{\bxi}_3,
\widehat{\bxi}_4$ are the (right) eigenvectors of the matrix
\revt{$\A_n$}.  {We note that these eigenvectors are mutually
  orthogonal and that the matrices $\A_n$, $n \ge 1$, are normal.}  In
the light of ansatz \eqref{eq:zeta}, {the eigenvectors
  $\widehat{\bxi}_1, \widehat{\bxi}_2, \widehat{\bxi}_3,
  \widehat{\bxi}_4$} take the following forms in the ``physical''
space as functions of the coordinate $\gamma$
\begin{subequations}
\label{eq:xi}
\begin{align}
{\xi}_1(\gamma) & = \psi_n(\gamma)(1-i), \label{eq:xi1} \\ 
{\xi}_2(\gamma) & = \phi_n(\gamma)(1-i), \label{eq:xi2} \\ 
{\xi}_3(\gamma) & = \phi_n(\gamma)(1+i), \label{eq:xi3} \\ 
{\xi}_4(\gamma) & = \psi_n(\gamma)(1+i), \label{eq:xi4}
\end{align}
\end{subequations}
where
\begin{equation}
\phi_n(\gamma) = \cos(n\gamma) - \sin(n\gamma), \qquad
\psi_n(\gamma) = \cos(n\gamma) + \sin(n\gamma).
\label{eq:phipsi}
\end{equation}
These functions satisfy the relation $\phi_n(\gamma) = \psi_n(2\pi -
\gamma)$ which implies that the pairs of unstable and stable
eigenvectors, respectively, $\xi_{1/2}$ and $\xi_{3/4}$, have the same
form, but different phases.  The equilibrium configurations
$\tilde{z}$ perturbed with the unstable eigenvectors $\xi_1$ and
$\xi_2$ \revt{defined for a single wavenumber $n=1$,
  cf.~\eqref{eq:zpert}, and with perturbations obtained by superposing
  components $\xi_1$ and $\xi_2$ with different wavenumbers
  $n=1,2,3,4$} are shown in the ``physical'' space in figure
\ref{fig:xi}.  Noting that
\begin{equation*}
\cos(n\gamma) = \frac{1}{2}\left[\psi_n(\gamma) + \phi_n(\gamma)\right], \quad
\sin(n\gamma) = \frac{1}{2}\left[\psi_n(\gamma) - \phi_n(\gamma)\right],
\end{equation*}
we conclude that every perturbation $\hzeta_n(t) \mbox{e}^{i n
  \gamma}$ with the wavenumber $n$ can be uniquely decomposed in terms
of the four eigenvectors \eqref{eq:xi1}--\eqref{eq:xi4}. The above
observations will simplify our analysis of the control-theoretic
properties of the system presented in \S \ref{sec:control}. 

We thus conclude that the linearization of the Birkhoff-Rott equation
in the neighbourhood of the flat-sheet equilibrium $\tilde{z}$ is
characterized by a discrete spectrum of orthogonal unstable modes
whose growth rates increase in proportion to the wavenumber $n$. We
add that regularization approaches mentioned in \S \ref{sec:intro}
would strongly suppress this instability at higher wavenumbers by
introducing an upper bound on the growth rates \citep{hnp06a}.

\subsection{Actuation}
\label{sec:actuation}

The form of actuation used in our \revt{study} is an idealized
representation of the forcing typically employed in actual experiments
involving control of shear layers, i.e., distributed blowing and
suction, microjets, etc. \revt{To account for such actuation, we
  modify the model introduced above by incorporating two hypothetical
  horizontal solid boundaries located a certain distance $b_0 > 0$
  above and below the sheet in its undeformed configuration (figure
  \ref{fig:setting}). We then suppose that actuation is produced by
  arrays of point sink/sources or point vortices located at each of
  the boundaries. We note that in the absence of actuation and when
  the vortex sheet is in its equilibrium (undeformed) configuration
  $\tilde{z}$, the corresponding flow induced by the sheet is
  everywhere parallel to the boundaries, hence it satisfies the
  no-through-flow boundary condition required by the Euler equation.
  On the other hand, when the sheet is deformed as in
  \eqref{eq:zpert}, a small wall-normal velocity, of order
  $\O(\varepsilon / b_0)$, is induced on the boundaries.  In
  principle, it could be eliminated by including a suitable potential
  velocity field in the model \eqref{eq:BR}, however, in order to keep
  our model analytically tractable, we neglect this effect.}

\begin{figure}
\centering
\mbox{
\subfigure[]{\includegraphics[width=0.485\textwidth]{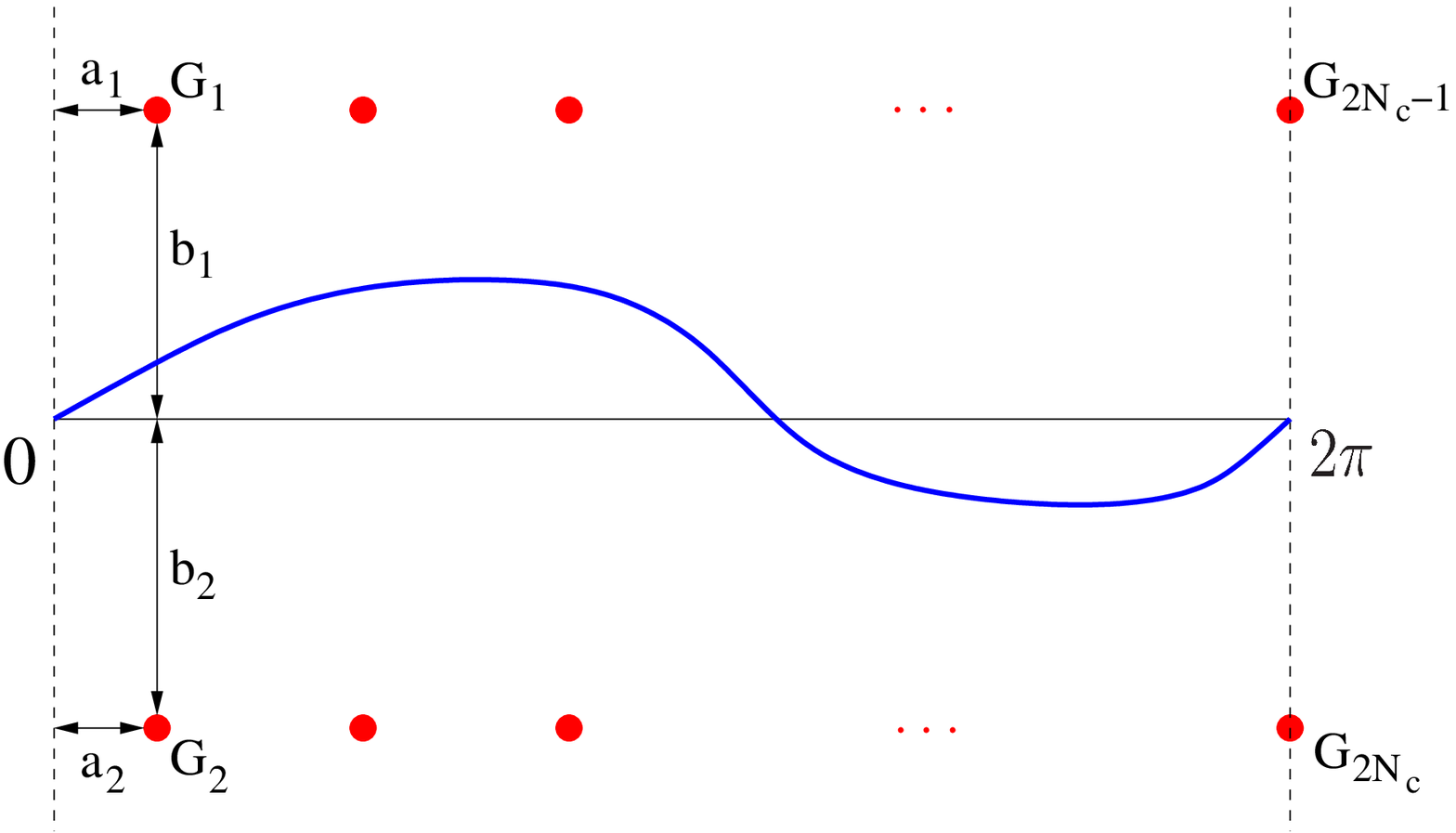}}
\qquad
\subfigure[]{\includegraphics[width=0.485\textwidth]{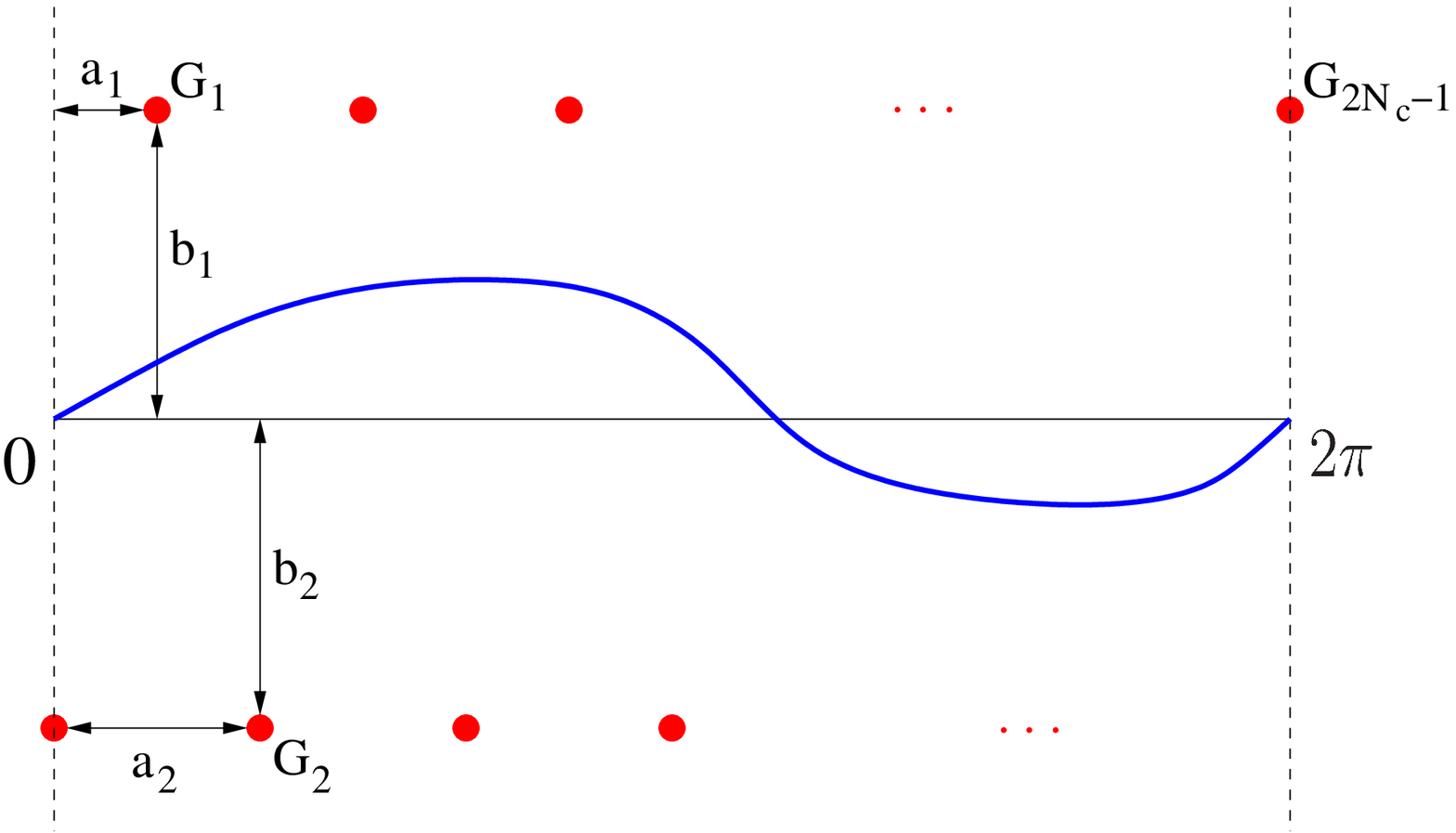}}}
\caption{Schematic representations of the actuation set-up with an (a)
  aligned and (b) staggered arrangement of point vortices or
  sinks/sources (denoted by red solid circles). The undeformed and
  perturbed vortex sheets are represented with, respectively, thin
  black and thick blue lines.}
\label{fig:actuators} 
\end{figure}

Each of the \revt{boundaries contains an array of $N_c$ point
  sinks/sources or point vortices}, so that the total number of
actuators is $2N_c$.  Their locations will be denoted $w_k = a_k +
ib_k$, where $a_k,b_k \in \RR$, $k=1,\dots,2N_c$, with odd (even)
indices {$k$} corresponding to actuators above (below) the sheet.
Since the two \revt{boundaries} are located at the same distance
\revt{$b_0$} from the sheet, we have $b_1 = b_3 = \ldots = b_{2N_c -
  1} = b_0$ and $b_2 = b_4 = \ldots = b_{2N_c} = -b_0$. Since each of
the arrays consists of actuators equi-spaced in the horizontal
direction, we also have $a_{2k+1} = a_1 + 2\pi k / N_c$ and $a_{2k+2}
= a_2 + 2\pi k / N_c$, $k=1,\dots,N_c-1$. As regards the relative
location of the actuators in the two arrays, we will consider two
arrangements: aligned (with $a_1 = a_2$, see figure
\ref{fig:actuators}(a)) and staggered (with $|a_2 - a_1| = \pi / N_c$,
see figure \ref{fig:actuators}(b)). We note that since due to
periodicity the system is invariant with respect to translation in the
$x$ direction, the actual values of $a_1$ and $a_2$ are unimportant
and what matters is {only} the difference $|a_2 - a_1|$. The velocity
induced at \revt{a certain point $\chi \in \CC$} by an individual, say
$k$-th, actuator is given by the expression $(G_k / (4 \pi i))
\cot\left((\revt{\chi}-w_k)/2 \right)$, \revt{$\chi \neq w_k$}, where
$G_k \in \CC$ is the strength of the actuator. It \revt{can} be given
{\em either} in terms of the \revt{intensity $Q_k \in \RR$ of the
  point sink/source, i.e., $G_k = iQ_k$, {\em or} in terms of the
  circulation $\Gamma_k \in \RR$ of the point vortex, i.e., $G_k =
  \Gamma_k$}. \revt{As regards the two forms of the actuation, we
  remark that point sinks/sources, which can be viewed as idealized
  models for localized blowing-and-suction, generate velocity fields
  which are radial with respect to the actuator location $w_k$.
  Therefore, such a velocity field does not violate the
  no-through-flow boundary condition on the horizontal wall on which
  the actuator is mounted (this boundary condition is in fact violated
  by the velocity field induced by the actuators located at the
  opposite boundary, regardless of their form, but this effect is of
  order $\O(b_0^{-1})$ and could be therefore neglected \revt{unless
    the actuators are placed close to the vortex sheet}). On the
  other hand, point vortices, which can be viewed as idealized models
  for localized vorticity generation, induce velocity fields with
  closed streamlines concentric around the actuator location $w_k$.
  Therefore, this velocity field will significantly violate the
  no-through-flow condition on the horizontal boundary in the
  neighbourhood of the actuator location $w_k$. For this reason, in
  this study we will primarily focus on point sinks/sources used as
  actuators and, for the sake of completeness, we will also briefly
  consider actuation with point vortices.}

The time-dependent \revt{sink/source intensities and vortex
  circulations, $Q_k = Q_k(t)$ and $\Gamma_k = \Gamma_k(t)$}, will be
determined by a feedback-control algorithm which will be designed in
\S \ref{sec:LQR}.  In the presence of such actuation, the
Birkhoff-Rott equation \eqref{eq:BR} takes the following form:
\begin{equation}
\frac{\partial z^\ast}{\partial t}(\gamma,t) = 
\frac{1}{4\pi i}\mbox{pv} \int_0^{2\pi} \cot\left( \frac{z(\gamma,t)-z(\gamma^\prime,t) }{2} \right)\,d\gamma^\prime + 
\sum_{k=1}^{2N_c} \frac{G_k}{4 \pi i} \cot\left(\frac{z(\gamma,t)-w_k}{2} \right).
\label{eq:BRc}
\end{equation}

\revt{The actuator strengths $G_1(t),\dots,G_{2N_c}(t)$ are not all
  independent. When point sinks/sources are used, mass must be
  conserved in time not only globally, but also independently in each
  of the subdomains above and below the sheet, as otherwise there
  would be a net mass flux across the sheet, which is undesirable.
  Consequently, the actuation must act such that the net mass flux
  produced by the point sinks/sources both above and below the sheet
  is zero. Likewise, when point vortices are used, their total
  circulation must also be conserved in order to satisfy Kelvin's
  theorem. Moreover, to ensure that such actuation does not generate a
  net flow in the horizontal direction, the circulations of the point
  vortices above and below the sheet must independently add up to
  zero. These constraints can be expressed as the following relations
\begin{subequations}
\label{eq:const}
\begin{alignat}{4}
&G_1(t)&+&G_3(t)&+\dots+&G_{2Nc-1}(t)& &= 0, \label{eq:const1} \\
&G_2(t)&+&G_4(t)&+\dots+&G_{2Nc}(t)& &= 0 \label{eq:const2}
\end{alignat}
\end{subequations}
which must be satisfied by the actuator strengths at all times.}

To summarize, the choices determining the actuation set-up are as
follows: (i) the number $N_c$ of actuator pairs, (ii) aligned versus
staggered arrangement of the actuators and (iii) the parameter $b_0$
representing the distance between the actuators and the undeformed
vortex sheet.

\section{Control-Oriented Characterization of the System}
\label{sec:control}

In this section we first characterize the input-to-state properties of
the linearized system \eqref{eq:BRlin} from the control-theoretic
point of view. This will allow us to assess whether the
Kelvin-Helmholtz instability analyzed in \S \ref{sec:KH} can be
stabilized using a linear state-feedback control approach based on the
actuation described in \S {\ref{sec:actuation}}. Then, based on these
results, we will design an LQR stabilization approach.

\revt{From the discussion at the end of \S \ref{sec:actuation},
  cf.~\eqref{eq:const}, it is evident that at least \revt{one
    actuator pair must be used both above and below the vortex sheet,}
  and we will therefore require $N_c \ge 2$.  To simplify analysis, it
  will be more convenient to impose constraints
  \eqref{eq:const1}--\eqref{eq:const2} after deriving the canonical
  representation of the controlled system.}  The controlled
Birkhoff-Rott equation \eqref{eq:BRc} \revt{can be rewritten in the
  following standard form}
\begin{equation}
\frac{\partial z^\ast}{\partial t} = V(z)+ \revt{\tb(z)\G,}
\label{eq:BRc2}
\end{equation}
where $V(z)$ is defined in \eqref{eq:BR}, \revt{$\G = [ G_1, \ \dots \ ,G_{2N_c} ]^T$} and 
\begin{align}
\revt{\tb(z)} = &  \revt{\left[ \begin{array}{ccc} C_1(z), & \dots &, C_{2N_c}(z)  \end{array}\right]} \label{eq:b} \\
& \text{with} \quad
C_k(z) =  \frac{1}{4\pi i} \cot\left( \frac{z-a_k-ib_k}{2} \right), \quad \revt{k = 1,\dots,2N_c}.
\nonumber
\end{align}
\revt{We use the convention that tilde ($\widetilde{\phantom{\_}}$)
  denotes control matrices before application of constraints
  \eqref{eq:const1}--\eqref{eq:const2}.}  Linearizing equation
\eqref{eq:BRc2} around the flat-sheet configuration as described in \S
\ref{sec:KH}, with the control input represented as \revt{$\G =
  \varepsilon \g = \varepsilon [g_1, \ \dots \ ,g_{2N_c}]^T$,} where
\revt{$g_1, \dots, g_{2N_c} \in \CC$}, we obtain
\begin{equation}
\frac{\partial \zeta^\ast}{\partial t} = V^\prime(\gamma)\zeta + \revt{\tb(\gamma) \g},
\label{eq:dzdtB}
\end{equation}
in which $V^\prime(\gamma)\zeta$ denotes the linearization of the
Birkhoff-Rott integral given in \eqref{eq:BRlin} and we note that the
term involving \revt{$\tb^\prime(\gamma)\zeta$} is absent, because the control
input \revt{${\G}$} vanishes in the equilibrium configuration. Using
relations \eqref{F-cot-p} and \eqref{F-cot-n} from Appendix
\ref{sec:spectral}, the components of the vector \revt{$\tb(\gamma)= \left[
  \begin{array}{ccc} C_1(\gamma), & \dots &, C_{2N_c}(\gamma)
  \end{array} \right]$} can be represented as follows, 
\revt{cf. relation \eqref{eq:b} with $z=\gamma$},
\begin{subequations}
\label{eq:C}
\revt{\begin{alignat}{2}
C_k(\gamma) &= \frac{1}{4\pi} +\frac{1}{2\pi} \sum_{n=1}^\infty \mbox{e}^{-n  b_k + i n a_k}\, \mbox{e}^{-in\gamma},& 
\quad k &= 1,3,\dots,2N_c-1,  \label{eq:C1} \\
C_k(\gamma) &=  - \frac{1}{4\pi}  - \frac{1}{2\pi} \sum_{n=1}^\infty \mbox{e}^{n b_k - i n a_k}\, \mbox{e}^{in\gamma},& 
\quad k &= 2,4,\dots,2N_c. \label{eq:C2}
\end{alignat}}
\end{subequations}
Substituting ansatz \eqref{eq:zeta} into equation \eqref{eq:dzdtB} and
taking note of expansions \eqref{eq:C}, we obtain the spectral
representation of the linear system \eqref{eq:dzdtB} which remains
{block-diagonal}, cf.~\eqref{Leq}, with the Fourier coefficients
$\{\hzeta_{-n}^\ast, \hzeta_n\}$, $n \ge 1$, appearing as the state
variables
\begin{equation}
\begin{aligned}
& \frac{\mbox{d}}{\mbox{d}t}\left[ \begin{array}{c} \zeta_{-n}^\ast \\ \zeta_n \end{array} \right] = 
\left[ \begin{array}{cc} 0 & \frac{i n}{2} \\  -\frac{i n}{2} & 0 \end{array}\right]\left[ \begin{array}{c} \zeta_{-n}^\ast \\ \zeta_n \end{array} \right] \\
& + \revt{\frac{1}{2\pi}\left[ 
\begin{array}{ccccc} 0 &  - \mbox{e}^{n b_2 -i n a_2} & \cdots & 0 &  - \mbox{e}^{n b_{2N_c} -i n a_{2N_c}} \\  
\mbox{e}^{-n b_1 - i n a_1} & 0 & \cdots & \mbox{e}^{-n b_{2N_c-1} - i n a_{2N_c-1}} & 0 \end{array}\right] 
\left[\begin{array}{c} g_1 \\ g_{2} \\ \vdots \\ g_{2N_c-1} \\  g_{2N_c} \end{array} \right]}.
\end{aligned}
\label{eq:dzetadt}
\end{equation}
The corresponding real-valued form of system \eqref{eq:dzetadt},
cf.~\eqref{eq:A}, will depend on whether the actuation \revt{$\G$}
involves \revt{point sinks/sources or point vortices}, cf.~\S
\ref{sec:actuation}, and the two cases are discussed below.

\subsection{Control with \revt{Point Sinks/Sources}}
\label{sec:control_ss}

We begin by considering actuation with point \revt{sinks/sources},
i.e., by setting $g_k = \revt{iQ_k}$, \revt{$k=1,\dots,2N_c$}, in
\eqref{eq:dzetadt}. Rewriting system \eqref{eq:dzetadt} in terms of
the state vector of real-valued coefficients, cf.~\eqref{eq:A}, we
obtain for each wavenumber $n$
\begin{equation}
\frac{\mbox{d}{\bm X}_n}{\mbox{d}t} = 
\A_n {\bm X}_n + \revt{\tB_n^{Q} \,\g}, \quad n \ge 1,
\label{eq:ABG0}
\end{equation}
where \revt{$\tB_n^{Q} = \left[ \B_{n,1}^Q, \ \dots \ ,\B_{n,N_c}^Q
  \right]$ in which
\begin{equation}
\B_{n,k}^Q = 
\left[ \begin{array}{cc} 
0 & \frac{1}{2\pi}\mbox{e}^{ n  b_{2k}}\sin(  n  a_{2k}) \\ 
-\frac{1}{2\pi} \mbox{e}^{- n  b_{2k-1}} \cos(  n  a_{2k-1}) & 0 \\
0 & -\frac{1}{2\pi} \mbox{e}^{ n  b_{2k}} \cos(  n  a_{2k}) \\ 
-\frac{1}{2\pi} \mbox{e}^{- n  b_{2k-1}} \sin(  n  a_{2k-1}) & 0
\end{array} \right], \quad k=1,\dots,N_c
\label{q:BQk}
\end{equation}
represents the block corresponding to the $k$th pair of point
sinks/sources.} \revt{In principle, constraints
\eqref{eq:const1}--\eqref{eq:const2} could be used to eliminate two
control variables, each associated with actuators on one side of the
vortex sheet, for example, $Q_{2N_c-1} = - Q_1 - \ldots - Q_{2N_c-3}$
and $Q_{2N_c} = - Q_2 - \ldots- Q_{2N_c-2}$. However, such an approach
would be undesirable from the computational point of view, because
certain actuators (namely, the last two ones) would be treated
differently from the rest. An alternative solution which is
algebraically equivalent, but leads to a numerically better-behaved
problem, is to express the control input vector as
\begin{equation}
\revt{\hg} = \left[Q_1+\Phi, Q_2+\Psi,\ \dots \ , Q_{2N_c-1}+\Phi, Q_{2N_c}+\Psi \right]^T,
\label{eq:gc}
\end{equation}
where $\Phi,\Psi \in \RR$ can be viewed as ``Lagrange multipliers''
offering two additional degrees of freedom necessary to accommodate
constraints \eqref{eq:const1}--\eqref{eq:const2}. Using these
constraints to eliminate $\Phi$ and $\Psi$ we obtain from \eqref{eq:gc}
\begin{equation}
\revt{\hg} = \begin{bmatrix*}[l]
Q_1 - \frac{1}{N_c} \left( Q_1+Q_3+\dots+Q_{2N_c-1}\right) \\
Q_2 - \frac{1}{N_c} \left( Q_2+Q_4+\dots+Q_{2N_c}\right) \\
\phantom{Q_2} \dots \\
Q_{2N_c-1} - \frac{1}{N_c} \left( Q_1+Q_3+\dots+Q_{2N_c-1}\right) \\
Q_{2N_c} - \frac{1}{N_c} \left( Q_2+Q_4+\dots+Q_{2N_c}\right) 
\end{bmatrix*},
\label{eq:g2}
\end{equation}
such that after replacing $\g$ in \eqref{eq:ABG0} with \revt{$\hg$}
given in \eqref{eq:g2} and rearranging terms we can define the control
matrix corresponding to the constrained actuation as
\begin{equation}
\begin{aligned}
\B_n^{Q} = \bigg[ & \B_{n,1}^{Q} - \frac{1}{N_c} \left( \B_{n,1}^{Q}+\B_{n,3}^{Q}+\dots+\B_{n,2N_c-1}^{Q}\right), \\
           & \B_{n,2}^{Q} - \frac{1}{N_c} \left( \B_{n,2}^{Q}+\B_{n,4}^{Q}+\dots+\B_{n,2N_c}^{Q} \right), \\
           & \phantom{Q_2} \dots \\
           & \B_{n,2N_c-1}^{Q} - \frac{1}{N_c} \left( \B_{n,1}^{Q}+\B_{n,3}^{Q}+\dots+\B_{n,2N_c-1}^{Q}\right), \\
           & \B_{n,2N_c}^{Q} - \frac{1}{N_c} \left( \B_{n,2}^{Q}+\B_{n,4}^{Q}+\dots+\B_{n,2N_c}^{Q} \right)  \bigg].
\end{aligned}
\label{eq:BnQ}
\end{equation}
}\revt{Thus, in the presence of constraints
  \eqref{eq:const1}--\eqref{eq:const2}, at every wavenumber $n$ the
  linearized problem takes the following form
\begin{equation}
\frac{\mbox{d}{\bm X}_n}{\mbox{d}t} = 
\A_n {\bm X}_n + \B_n^{Q}\, \g, \quad n \ge 1.
\label{eq:ABG}
\end{equation}
}\revt{We remark that while the entries of the control input vector
  $\g = \left[Q_1, \ \dots \, Q_{2N_c}\right]^T$ need not satisfy
  constraints \eqref{eq:const1}--\eqref{eq:const2}, the structure of
  the control matrix $\B_n^{Q}$, cf.~\eqref{eq:BnQ}, ensures that the
  constraints are in fact satisfied by the actuation \revt{velocity
    field} represented by the terms $\B_n^{Q}\g$, $n \ge 1$.}

Defining \revt{the vectors $\X = \left[ \X_1^T,\ldots,\X_N^T
  \right]^T$ and $\B^{Q} =
  \left[\left(\B_1^{Q}\right)^T,\ldots,\left(\B_N^{Q}\right)^T\right]^T$
  and the matrix $\A = \diag\left(\A_1,\ldots,\A_N\right)$} for some
{$N \ge 1$}, the evolution of system \eqref{eq:ABG} truncated for $n
\le N$ is described by
\begin{equation}
 \frac{\mbox{d}\X}{\mbox{d}t} =  \A \X + \revt{\B^{Q}}\, \g.
\label{eq:ABG_N}
\end{equation}
{For a system in this form the property of controllability means
  that for all initial conditions $\X_0$ the control can always drive
  the solution $\X(t)$ to zero. It can be established} by verifying
whether the following rank condition is satisfied \citep{s94}
\begin{equation}
\mbox{Rank}\left[\revt{\B^Q \ \A\B^Q \ \ldots \ \A^{N-1}\B^Q} \right] 
= 4N.
\label{eq:RG}
\end{equation}
We observe that, due to the block-diagonal structure of the matrix
$\A$, a necessary and sufficient condition for relation \eqref{eq:RG}
to hold for $N \rightarrow \infty$ is that the corresponding rank
conditions for individual blocks be satisfied simultaneously, i.e.,
\begin{equation}
\R_n^Q = \mbox{Rank}\left[\revt{\B_n^Q \ \A_n\B_n^Q \ \A_n^2 \B_n^Q \  \A_n^3 \B_n^Q} \right]=4, \quad n \ge 1.
\label{eq:RGn}
\end{equation}
It can be verified using symbolic-algebra tools such as {\tt Maple}
that condition \eqref{eq:RGn} does indeed hold for $n \ge 1$ and for
almost all values of the parameters $a_1, a_2, b_1$ and $b_2$,
\revt{provided
\begin{equation}
n \neq \ell N_c \quad \text{for any integer} \ \ell.
\label{eq:nNc}
\end{equation}
This implies that matrix pairs $\{\A_n,\B_n^Q\}$ are controllable for
all $n=1,\dots,N$ if the number $N_c$ of actuator pairs is larger than
the largest wavenumber $N$ present in system \eqref{eq:ABG_N}.
Evidently, the simplest way to satisfy condition \eqref{eq:nNc} is to
set $N_c = N+1$.  In particular, in the continuous limit corresponding
to $N \rightarrow \infty$, we conclude that the linearized
Birkhoff-Rott equation is controllable when the actuation has the form
of a continuous distribution of sinks/sources. When condition
\eqref{eq:nNc} is not satisfied, then $\R_n^Q = 0$, implying that both
unstable eigenmodes with wavenumber $n$ are uncontrollable. We add
that condition \eqref{eq:nNc} is a consequence of constraints
\eqref{eq:const1}--\eqref{eq:const2}, as in their absence matrix pairs
$\{\A_n,\B_n^Q\}$ are controllable with $N_c = 1$ for all $n \ge 1$.}

It is interesting to understand how the control authority of the point
\revt{sinks/sources} depends on their locations represented by the
parameters $a_1$, $a_2$, $b_1$ and $b_2$ (cf.~\revt{figures
  \ref{fig:actuators}(a) and \ref{fig:actuators}(b)}). To assess this
we will analyze the controllability residuals $\bsigma_1$ and
$\bsigma_2$ of the unstable modes $\widehat{\bxi}_1$ and
$\widehat{\bxi}_2$, cf.~\eqref{eq:hxi12}. They quantify the authority
the control actuation has over individual modes and, \revt{for
  clarity, we will restrict our attention here to the case with
  $N_c=2$ so that we will only consider the wavenumber $n=1$,
  cf.~\eqref{eq:nNc}} (to simplify the notation, we will not indicate
\revt{this} wavenumber on the different variables defined in the
remainder of this subsection).  We expand the state $\Xn$ in terms of
eigenvectors \eqref{eq:hxi12}--\eqref{eq:hxi34} as $\Xn(t) =
\sum_{k=1}^4 {\rho_k}(t) \widehat{\bxi}_k$.  Substituting this
expansion into \eqref{eq:ABG} and taking the inner product of this
equation and the left eigenvector {$\widehat{\bbeta}_l$} of the matrix
$\A_n$ we obtain
\begin{equation}
 \frac{\mbox{d}\rho_l}{\mbox{d}t} = \lambda_l \rho_l + \bsigma_l \g, \qquad l=1,2,3,4,
\label{eq:rho}
\end{equation}
where $\lambda_1 = \lambda_2 = \lambda_n^+$, $\lambda_3 = \lambda_4 =
\lambda_n^-$, cf.~\S \ref{sec:KH}, and we used the property of the
left and right eigenvectors $\widehat{\bbeta}_l^T \widehat{\bxi}_k =
\delta_{lk}$. We note that in the present problem the matrix $\A_n$ is
symmetric and hence the left and right eigenvectors coincide
$\widehat{\bbeta}_k = \widehat{\bxi}_k$, $k=1,2,3,4$. The
controllability residuals corresponding to the two unstable eigenmodes
$\widehat{\bxi}_1$ and $\widehat{\bxi}_2$ are
\begin{subequations}
\label{eq:sigma}
\revt{\begin{align}
\bsigma_1 & = \frac{1}{2\pi} \left[ \mbox{e}^{-b_1} (\cos(a_1)-\cos(a_1+\pi)) \quad 
  \mbox{e}^{b_2} (\sin(a_2)-\sin(a_2+\pi)) \right], \label{eq:sigma1} \\
\bsigma_2 & = \frac{1}{2\pi} \left[ \mbox{e}^{-b_1} (\sin(a_1)-\sin(a_1+\pi)) \quad 
  \mbox{e}^{b_2} (\cos(a_2)-\cos(a_2+\pi)) \right]. \label{eq:sigma2}
\end{align}}
\end{subequations}
\revt{Inspection of the entries of the vectors $\bsigma_1$ and
  $\bsigma_2$ reveals that they can {\em simultaneously} vanish for
  certain actuator locations, namely, for $a_1 = \pi/2$ and $a_2 = 0$
  in $\bsigma_1$ and for $a_1 = 0$ and $a_2 = \pi/2$ in $\bsigma_2$,
  which coincide with their staggered arrangement (cf.~figure
  \ref{fig:actuators}(b)). In fact, a similar pattern can also be
  observed for \revt{$N_c \ge 2$} which indicates that while system
  \eqref{eq:ABG_N} is controllable subject to condition \eqref{eq:nNc}
  for generic values of the parameters $a_1$ and $a_2$, the staggered
  configuration illustrated in figure \ref{fig:actuators}(b)
  represents in fact a degenerate case in which controllability is
  lost.  Although controllability in the staggered arrangement can in
  principle be re-established by taking a larger than minimal number
  of actuator pairs $N_c > N+1$, we will not further consider this
  configuration \revt{and will instead focus on the aligned
    arrangement, cf.~figure \ref{fig:actuators}(a), which has better
    controllability properties}.

  As regards the effect of the distance $b_0$ between the actuators
  and the vortex sheet on the control residuals \eqref{eq:sigma}, we
  remark that for larger numbers \revt{$N_c \ge 2$} of actuator pairs
  the factors $\mbox{e}^{-b_1}$ and $\mbox{e}^{b_2}$ in
  \eqref{eq:sigma} are replaced with $\mbox{e}^{-n b_1}$ and
  $\mbox{e}^{n b_2}$, where $n=1,\dots,N_c-1$. We then observe} that
the magnitudes of the controllability residuals do not achieve maxima
for $b_1 > 0$ and $b_2 < 0$, and vanish exponentially as the control
vortices move away from the vortex sheet, i.e., as $b_1 \rightarrow
\infty$ and $b_2 \rightarrow - \infty$ with the decay rate given by
the wavenumber $n$.  This implies that, for fixed locations
\revt{$(a_k,b_k)$, $k=1,\dots,2N_c$, the sinks/sources have control
authority} which decreases exponentially with the wavenumber $n$ of the
perturbation.  We emphasize that, at the same time, for increasing
wavenumbers $n$ the perturbations also become more unstable, cf.~\S
\ref{sec:KH}. It is thus evident that, as the wavenumber $n$ gets
larger, stabilization of the corresponding unstable modes becomes more
difficult, because of the widening disparity between the growth rate
and the control authority which change in proportion to, respectively,
$n$ and $e^{-n}$. As will be discussed in \S \ref{sec:numer},
combination of these two factors will make the numerical computation
of the feedback {kernels} quite challenging.

\subsection{Control with \revt{Point Vortices}}
\label{sec:control_pv}

We now move on to \revt{briefly} consider actuation with point
\revt{vortices, i.e., by setting $g_k = \Gamma_k$, $k=1,\dots,2N_c$,
  in \eqref{eq:dzetadt}.  The analysis follows exactly the same lines
  and leads to qualitatively the same conclusions as in \S
  \ref{sec:control_ss} for actuation with point sinks/sources. In
  particular, by constructing the control matrices $\B_n^{\Gamma}$ as
  above with the blocks corresponding to the wavenumber $n$ and the
  $k$th point-vortex pair defined as
\begin{equation}
\B_{n,k}^{\Gamma} = 
\left[ \begin{array}{cc} 
0 &  -\frac{1}{2\pi}\mbox{e}^{ n  b_{2k}}\cos(  n  a_{2k}) \\ 
-\frac{1}{2\pi} \mbox{e}^{- n  b_{2k-1}} \sin(  n  a_{2k-1}) & 0 \\
0 & -\frac{1}{2\pi} \mbox{e}^{ n  b_{2k}} \sin(  n  a_{2k}) \\ 
\frac{1}{2\pi} \mbox{e}^{- n  b_{2k-1}} \cos(  n  a_{2k-1}) & 0 
\end{array} \right], \quad k=1,\dots,N_c,
\label{eq:BG}
\end{equation}
we conclude that matrix pairs $\{\A_n,\B_n^{\Gamma}\}$ are also
controllable for almost all values of the parameters $a_1$ and $a_2$
provided condition \eqref{eq:nNc} is satisfied \revt{(as was the case
  with point sink/source actuators in \S \ref{sec:control_ss},
  controllability is lost in the staggered configuration, cf.~figure
  \ref{fig:actuators}(b))}. Thus, system \eqref{eq:ABG_N} with $\B^Q$
replaced with $\B^{\Gamma}$ is controllable as long as sufficiently
many pairs of control vortices are used, i.e., $N_c > N$.}

\subsection{Formulation of the Stabilization Problem}
\label{sec:LQR}

We are now in the position to set up the LQR control problem for the
linearized Birkhoff-Rott equation \eqref{eq:dzdtB}. Since this form of
actuation \revt{is more justified on the physical grounds (cf.~the
  discussion in \S\ref{sec:actuation}), we will primarily} focus on
the control using point \revt{sinks/sources} as actuators.  \revt{In
  the light of the results about controllability from
  \S\ref{sec:control_ss}, cf.~condition \eqref{eq:nNc}, we will set
  the number of actuator pairs to $N_c \ge N+1$ and will only consider
  actuators in the aligned arrangement, cf.~figure
  \ref{fig:actuators}(a).} Our goal is thus to express the
\revt{intensities of sinks/sources $g_k = iQ_k$, $k=1,\dots,2N_c$,} as
functions of the instantaneous perturbation $\zeta(\gamma,t)$ of the
flat-sheet equilibrium, such that the resulting actuation, represented
by the second term in \eqref{eq:BRc} will drive the perturbation to
zero.  More precisely, we will be interested in eliminating only the
transverse component $y(\gamma,t) = \Im(\zeta(\gamma,t))$ of the
perturbation, i.e., we want to achieve $\| y(\cdot,t)\|_{\infty}
\rightarrow 0$ as $t \rightarrow \infty$, where $\| u \|_{\infty} =
\sup_{\gamma \in [0,2\pi]} | u(\gamma)|$.  Knowing that \revt{with the
  parameters chosen} the system is controllable, cf.~\S
\ref{sec:control_ss}, such a feedback control strategy can be designed
using an approach known as the Linear-Quadratic Regulator \citep{s94}.
In addition to stabilizing the equilibrium configuration, the feedback
operators determined in this way will also ensure that the resulting
transient system evolution minimizes a certain performance criterion
which will be defined below.  We add that, although this control
problem is designed in a purely deterministic setting, the LQR control
design also remains optimal in the presence of stochastic disturbances
with Gaussian structure affecting the system evolution, a result known
as the ``separation principle''. In practical settings feedback
strategies relying on some limited system output rather than full
state information are more applicable. In such cases, feedback
controllers are usually combined with state estimators to form
``compensators'' and in the context of inviscid vortex control
problems such approaches were pursued by \citet{Protas2004pf,nps17}.
However, since the present study in the first place aims to assess the
fundamental opportunities and limitations inherent in the linear
feedback control of the inviscid Kelvin-Helmholtz instability, we will
focus here on the more basic problem with the full-state feedback.

{The LQR approach is applied in the discrete setting and as} a
discretization of the continuous linearized Birkhoff-Rott equation
\eqref{eq:dzdtB}, we consider its truncated Fourier representation
\eqref{eq:ABG_N} where the wavenumbers are restricted to the range
$n=1,\dots,N$, whereas the matrix \revt{$\B^{Q}$ now has the dimension
  $4N \times 2N_c$.}  Since the zero (mean) mode $\hzeta_0$ is
invariant under the uncontrolled evolution, cf.~\eqref{Leq3}, and the
control actuation may only affect its real part corresponding to the
horizontal displacement, cf.~\eqref{F-cot-p}--\eqref{F-cot-n}, there
is no need to include this mode in the control system.

The feedback loop will be closed by expressing the control vector $\g$
as a linear function of the state perturbation $\X$,
\begin{equation}
\g = - \K \X,
\label{eq:g}
\end{equation}
where $\K \in \revt{\RR^{2N_c \times 4N}}$ is the feedback
operator {(kernel)}.  It will be determined in order to stabilize
system \eqref{eq:ABG_N} while simultaneously minimizing the following
objective function
\begin{align}
& \J(\g) = \frac{1}{2} \int_0^\infty \int_0^{2\pi} 
\left[ y(\gamma,t)\right]^2 + r \,\left[ \b(\gamma)\g(t) \right]^2 \,d\gamma\,dt,
\label{eq:J} \\
&  \revt{
\begin{aligned} 
\text{where} \quad 
\b(\gamma) = \bigg[
  & C_1(\gamma)- \frac{1}{N_c} \left( C_1(\gamma) +  C_3(\gamma) + \dots +  C_{2N_c-1}(\gamma) \right), \\
  & C_2(\gamma)- \frac{1}{N_c} \left( C_2(\gamma) +  C_4(\gamma) + \dots +  C_{2N_c}(\gamma) \right), \\
  & \phantom{C_2(\gamma)} \dots \\
  & C_{2N_c-1}(\gamma)- \frac{1}{N_c} \left( C_1(\gamma) +  C_3(\gamma) + \dots +  C_{2N_c-1}(\gamma) \right), \\
  & C_{2N_c}(\gamma)- \frac{1}{N_c} \left( C_2(\gamma) +  C_4(\gamma) + \dots +  C_{2N_c}(\gamma) \right) 
\bigg]
\end{aligned}}
\nonumber
\end{align}
\revt{ensures that the actuation velocity satisfies constraints
  \eqref{eq:const1}--\eqref{eq:const2},} and the two terms measure,
respectively, the ``energy'' associated with the transverse
displacement of the deformed sheet and the ``work'' done by the
actuation, whereas $r > 0$ is a parameter characterizing the relative
``cost'' of the control. Hereafter, we will refer to the quantity $\|
y(t) \|_2^2 = \int_0^{2\pi} \left[ y(\gamma,t)\right]^2 \,d\gamma$ as
the ``perturbation energy''.  Under the assumption that the system is
truncated to the wavenumbers $n=1,\ldots,N$ and using the state
variables $\left\{\alpha_{-n}, \beta_n, \beta_{-n}, \alpha_n
\right\}_{n=1}^N$, cf.~\eqref{eq:dabdt}--\eqref{eq:A}, the objective
function can be expressed as
\begin{align}
& \J(\g) = \frac{1}{2} \int_0^\infty \X(t)^T \Q \X(t) + 
r \, \g^T (\revt{\B^{Q}})^T \revt{\B^{Q}} \g \, dt, \label{eq:J2} \\
& \text{where} \quad \Q = \frac{\pi}{2} \diag \left( 
\underbrace{\left[ 
\begin{array}{cccc} 1 & 0 & 0 & -1 \\  0 & 1 & 1 & 0 \\ 
0 & 1 & 1 & 0 \\ -1 & 0 & 0 & {1} \end{array}\right],
\ldots,
\left[ 
\begin{array}{cccc} 1 & 0 & 0 & -1 \\  0 & 1 & 1 & 0 \\ 
0 & 1 & 1 & 0 \\ -1 & 0 & 0 & {1} \end{array}\right]}_{N \ \text{blocks}}
\right)
\nonumber
\end{align}
is chosen such that only the transverse displacement \revt{of the
  vortex sheet} is taken into account. As one of the cornerstone
results of the linear control theory \citep{s94}, it is well known
that the feedback operator $\K$ can be expressed as
\begin{equation}
\K = \frac{1}{r} (\revt{\B^{Q}})^T \bP,
\label{eq:K}
\end{equation}
where the matrix $\bP$ is a $(4N \times 4N)$ symmetric
positive-definite solution of the algebraic Riccati equation
\begin{equation}
\A^T \bP + \bP \A + \Q - \frac{1}{r} \bP \revt{\B^{Q}} (\revt{\B^{Q}})^T \bP = {\bf 0}.
\label{eq:Ric}
\end{equation}
The resulting closed-loop system 
\begin{equation}
 \frac{\mbox{d}\X}{\mbox{d}t} =  (\A - \revt{\B^{Q}}\K) \X 
\label{eq:ABK_N}
\end{equation}
is then guaranteed to be exponentially stable for all initial data
$\X_0$. In addition to \eqref{eq:ABK_N}, we will also consider the
closed-loop version of the nonlinear Birkhoff-Rott equation utilizing
the same feedback operator $\K$, namely
\begin{equation}
\frac{\partial z^\ast}{\partial t} = V(z) - 
{\F}^{-1} \left(\revt{\B^{Q}}\K \, {\F} (z - \tilde{z}) \right),
\label{eq:BRc3}
\end{equation}
where ${\F} \; : \; L^2(0,2\pi;\CC) \rightarrow \RR^{4N}$ is a linear
operator mapping the state variables defined in the ``physical'' space
to their Fourier-space representation in terms of the coefficients
$\left\{\alpha_{-n}, \beta_n, \beta_{-n}, \alpha_n \right\}_{n=1}^N$.
\revt{When point vortices are used as actuation all steps are the
  same, except that the control matrix $\B^Q$ is replaced with
  $\B^{\Gamma}$, cf.~\eqref{eq:BG}.}  In the next section we discuss
the numerical determination of the feedback kernel $\K$ via solution
of the Riccati equation \eqref{eq:Ric} and the numerical integration
of the closed-loop systems \eqref{eq:ABK_N} and \eqref{eq:BRc3}
highlighting the various challenges involved in these tasks.

\section{Numerical Approach}
\label{sec:numer}

\begin{figure}
\centering
\mbox{
\subfigure[]{\includegraphics[width=0.5\textwidth]{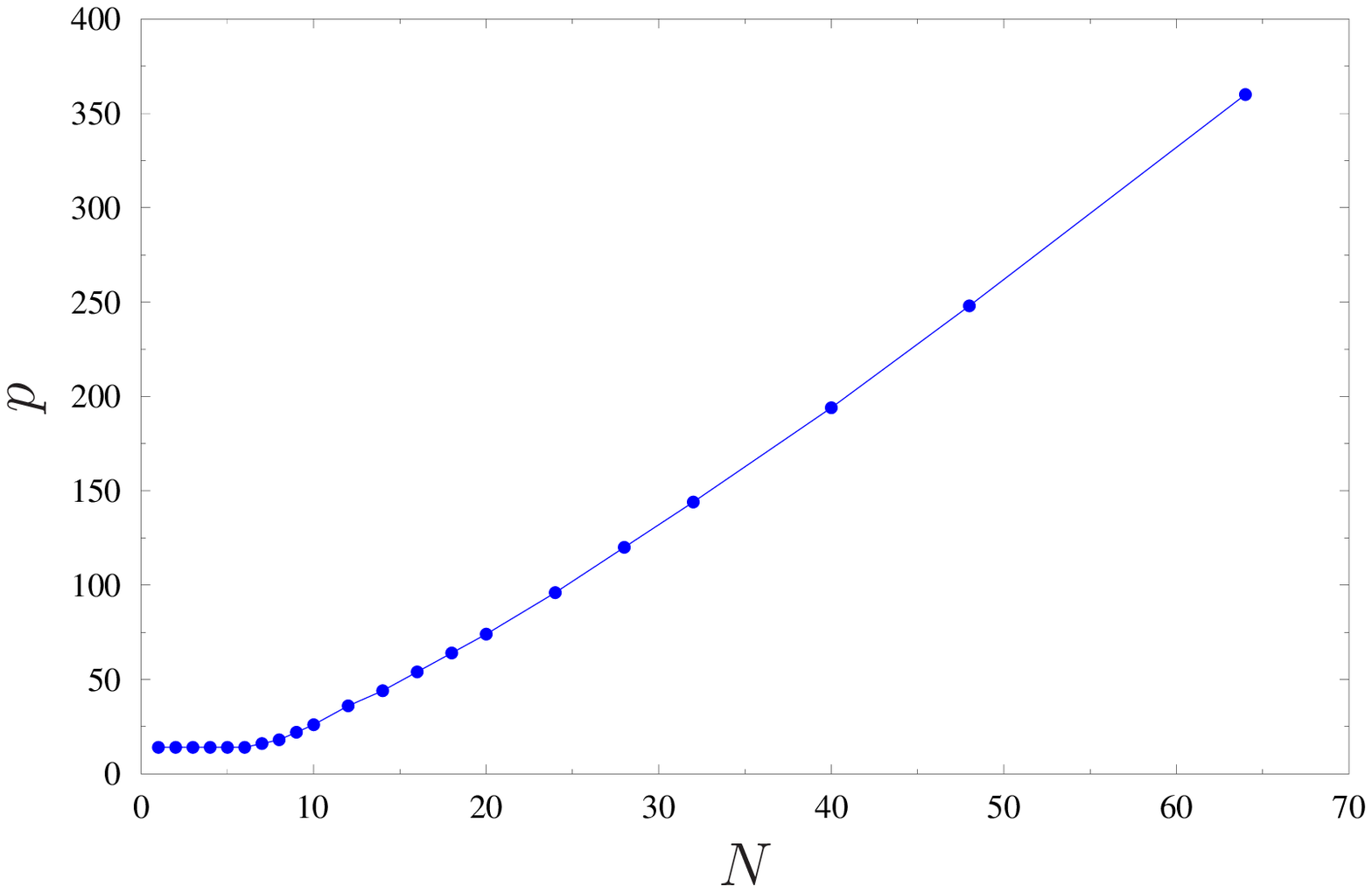}}
\quad
\subfigure[]{\includegraphics[width=0.5\textwidth]{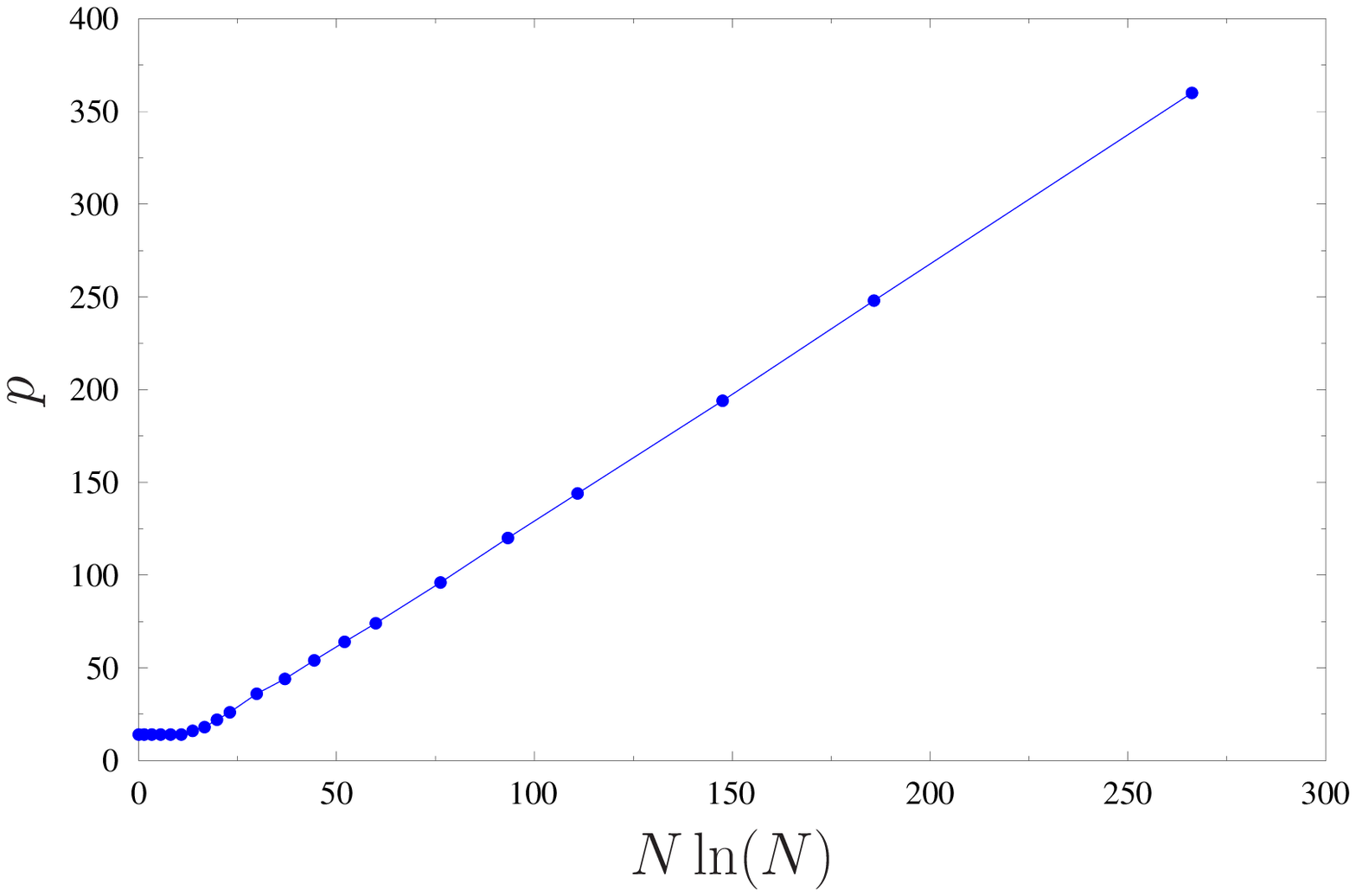}}}
\caption{Dependence of the minimum arithmetic precision $p$ required
  for an accurate solution of the Riccati equation \eqref{eq:Ric} and
  determination of the feedback kernels $\K$ on (a) {the
    resolution $N$ and on (b) the quantity} $N\ln(N)$. For ``large''
  $N$, a clear scaling $p \sim N\ln(N)$ is evident in panel (b).}
\label{fig:p} 
\end{figure}

In this section we first discuss the numerical solution of the Riccati
equation \eqref{eq:Ric} required in order to determine the feedback
operator $\K$ and then provide details about the numerical integration
of the linear and nonlinear closed-loop systems \eqref{eq:ABK_N} and
\eqref{eq:BRc3}. The first problem can in principle be solved using
the functions {\tt lqr} or {\tt care} from MATLAB's control toolbox
\citep{control_toolbox}. However, the main difficulty is that, as
discussed in \S \ref{sec:control_ss}, for increasing wavenumbers $n$
the growth rates of the unstable modes increase in proportion to $n$
while the authority which the actuators have over these modes
decreases in proportion to $e^{-n}$, resulting in a very poor
conditioning of the algebraic problem. Therefore, using the standard
implementations of the functions {\tt lqr} or {\tt care} based on the
\revt{double-precision arithmetics}, the Riccati equation can be
solved accurately only when system \eqref{eq:ABG_N} is truncated at a
very small wavenumber not exceeding $N = 6$ (we note that since every
block in \eqref{eq:ABG_N} contains 4 state variables
$\left\{\alpha_{-n}, \beta_n, \beta_{-n}, \alpha_n \right\}$, cf.~\S
\ref{sec:KH}, the actual state dimension is $4N$).  Problems related
to poor algebraic conditioning manifest themselves as the
impossibility to find a symmetric positive-definite solution $\bP$ of
\eqref{eq:Ric}, or, even if such a matrix $\bP$ can be found, by the
closed-loop system matrix $(\A - \revt{\B^{Q}}\K)$ still having unstable
modes. In order to overcome this difficulty, high-precision arithmetic
must be used to solve the Riccati equation \eqref{eq:Ric} and to
perform all other algebraic operations required to determine the
feedback kernel $\K$.  This is accomplished using {\tt Advanpix}, a
Multiprecision Computing toolbox for MATLAB \citep{Advanpix}, which
provides arbitrary-prevision versions of the functions {\tt lqr} and
{\tt care} in addition other functionality. For a given resolution
$N$, the minimum {arithmetic} precision, expressed in terms of the
number $p$ of significant digits needed to solve problem
\eqref{eq:Ric} and determine the feedback kernels, can be estimated by
evaluating numerically the rank of the composite matrix in
\eqref{eq:RG}. The result of this calculation will be less than $4N$,
unless a sufficient arithmetic precision is used. In the course of
extensive tests we determined that the smallest arithmetic precision
$p$ which ensures a correct numerical evaluation of the rank in
\eqref{eq:RG} will also be sufficient to accurately solve the Riccati
equation \eqref{eq:Ric} and determine the feedback kernel $\K$. The
dependence of thus determined minimum required arithmetic precision
$p$ on the resolution $N$ is shown in figure \ref{fig:p}. The change
of trend observed in this figure for small $N$ is due to the fact that
for such small resolutions the standard double precision
(corresponding to $p=14$) is sufficient. In figure \ref{fig:p}(b) we
see that for ``large'' $N$ the required precision $p$ grows in
proportion to $N\ln(N)$, with the approximate relationship $p \approx
1.3843 \, N\ln(N) - 9.1422$ obtained with linear regression. Using
this relation to extrapolate the data shown in figure \ref{fig:p}
indicates that to solve the problem with the resolution of $N = 128$,
arithmetic precision with about $p = 850$ significant digits would be
required (the RAM memory needed to solve the Riccati equation with
such values of $N$ and $p$ would exceed 128GB). The data presented in
figure \ref{fig:p} will guide our choice of the arithmetic precision
in computations with different resolutions $N$ discussed in \S
\ref{sec:results}.

In the numerical integration of the closed-loop systems
\eqref{eq:ABK_N} and \eqref{eq:BRc3} the same arithmetic precision $p$
is used as in the solution of the Riccati equation \eqref{eq:Ric} at
the given resolution $N$. The linear problem \eqref{eq:ABK_N} is
already an ODE system and is integrated using a multiprecision version
of MATLAB's adaptive time-stepping routine {\tt ode45} provided by
Advanpix \citep{Advanpix}. The relative and absolute tolerance are
both set to $10^{-10}$ which in the course of extensive tests was
found to ensure converged results.

The nonlinear closed-loop system \eqref{eq:BRc3} is discretized in
space using the collocation approach devised by \citet{vsheet:SaOk96},
but without regularization, on a {mesh} consisting of $2(N+1)$
equi-spaced grid points which ensures that solution components with
wavenumbers up to $N$ are resolved. The resulting ODE system is
integrated in time using the multiprecision {\tt ode45} routine
\citep{Advanpix}. The relative and absolute tolerance are both set to
$10^{-8}$ which in the course of extensive tests was found to ensure
converged results. We note that with sufficient arithmetic precision
there {is} no need to use the spectral-filtering technique introduced
by \citet{k86b} to control the spurious growth of round-off errors.
The operator $\F$ appearing in \eqref{eq:BRc3} is discretized using
discrete Fourier-transform techniques and, assuming that complex
numbers are represented as pairs of real numbers, the resulting matrix
$\bF$ has dimensions $4N \times 4(N+1)$, where the difference of the
two dimensions comes from the fact that the zero mode is not included
in the spectral representation, cf.~\S \ref{sec:LQR}. The matrix
$\bF^{-1}$ is then defined as a pseudo-inverse of $\bF$ with entries
given analytically in terms of the inverse discrete sine and cosine
transforms. A thorough assessment of the behavior of the linear and
nonlinear closed-loop systems \eqref{eq:ABK_N} and \eqref{eq:BRc3} in
function of several different parameters is offered in the next
section.

\FloatBarrier

\section{Results}
\label{sec:results}

In this section we present a number of computational results
illustrating the behavior of the linear and nonlinear closed-loop
systems \eqref{eq:ABK_N} and \eqref{eq:BRc3}, where we will analyze
the influence of the different parameters characterizing the
actuation, cf.~\S \ref{sec:actuation}. This will allow us to assess to
what degree the properties established theoretically for the linear
problem in \S \ref{sec:control_pv} can be extended to the nonlinear
setting.  \revt{Using point vortices, rather than point sinks/sources,
  as actuation results in essentially identical performance of the
  closed-loop system in the linear case and very similar performance
  in the nonlinear case. Therefore, for brevity, we will restrict the
  presentation of our results here to actuation with point
  sinks/sources.}  To fix attention, unless stated otherwise, we will
consider actuation with varying numbers $N_c$ of \revt{sink/source}
pairs in the aligned arrangement (cf.~figure \ref{fig:actuators}(a))
at the distance $b_0 = 1.0$ from the sheet in its undisturbed
configuration, whereas the relative cost of control appearing in
\eqref{eq:J} will be $r = 1.0$. In all cases the initial condition
{$z_0$} will be constructed by superposing a {suitably scaled
  perturbation $\zeta_0$ on the equilibrium configuration $\tilde{z}$,
  i.e., $z_0 = \tilde{z} + \varepsilon \, \zeta_0$, where the
  perturbation will be taken as the sum of two unstable eigenfunctions
  $\xi_1$ and $\xi_2$, cf.~\eqref{eq:xi1}--\eqref{eq:xi2},
\begin{equation}
\zeta_0 = \xi_1 + \xi_2
\label{eq:zeta0}
\end{equation}
with wavenumber $n_0$ which, unless stated otherwise, will be $n_0 =
1$, cf.~figure \ref{fig:xi}.}  The ``size'' of the initial
perturbation determined by {$\varepsilon$} will play an important role
in the nonlinear case, but is irrelevant in the linear setting. Due to
the reasons detailed in \S \ref{sec:numer}, the numerical resolution
used in most of the {computations} discussed below is rather modest
and given by \revt{$N = 31$}, corresponding to $N_x = 64$ grid points
in the ``physical'' space and the system dimension of 128. At this
resolution the required arithmetic precision is $p = 144$, cf.~figure
\ref{fig:p}. We recall that in the absence of control the
Birkhoff-Rott system \eqref{eq:BR} exhibits a finite-time blow-up
manifested by the appearance of a curvature singularity. \revt{In
  order to verify the analysis from \S \ref{sec:control_ss}, we first
  investigate the borderline case with the smallest possible number
  $N_c = N +1$ of actuator pairs and will then focus on the more
  relevant cases when \revt{$N_c \ge N +1$}.}

\begin{figure}
\centering
\includegraphics[width=0.5\textwidth]{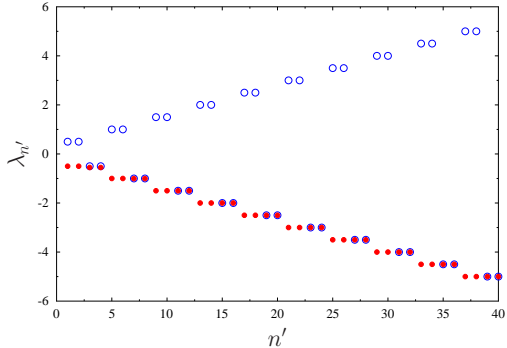}
\caption{Eigenvalues of the linear open-loop system \eqref{eq:A2}
  (empty blue circles) and of the closed-loop system \eqref{eq:ABK_N}
  (filled red circles) for the resolution \revt{$N = 31$} and
  actuation with \revt{$N_c = N + 1 = 32$} pairs of
  \revt{sinks/sources}. The index $n'$ is defined such that $n = (n'
  \mod 4)$ and, for clarity, only eigenvalues with small magnitude are
  shown.}
\label{fig:evals} 
\end{figure}

To begin, we show the eigenvalues of the linear open-loop and the
closed-loop systems \eqref{eq:A2} and \eqref{eq:ABK_N} in figure
\ref{fig:evals}. In the uncontrolled case at every wavenumber $n$
there are two positive and two negative eigenvalues $\lambda_n^+$ and
$\lambda_n^-$, cf.~\S \ref{sec:KH}. In the controlled case the
originally unstable eigenvalues $\lambda_n^+$ have their signs
reversed with magnitude approximately retained. The matrix $(\A -
\B^{Q}\K)$ of the {linear closed-loop} system is not
block-diagonal and normal, in contrast to the matrix $\A$ of the
open-loop system, cf.~\S \ref{sec:KH}. \revt{However, the degree of
  its non-normality is fairly weak and decreases as the number $N_c$
  of actuator pairs increases.}

\begin{figure}
\centering
\mbox{
\subfigure[]{\includegraphics[width=0.5\textwidth]{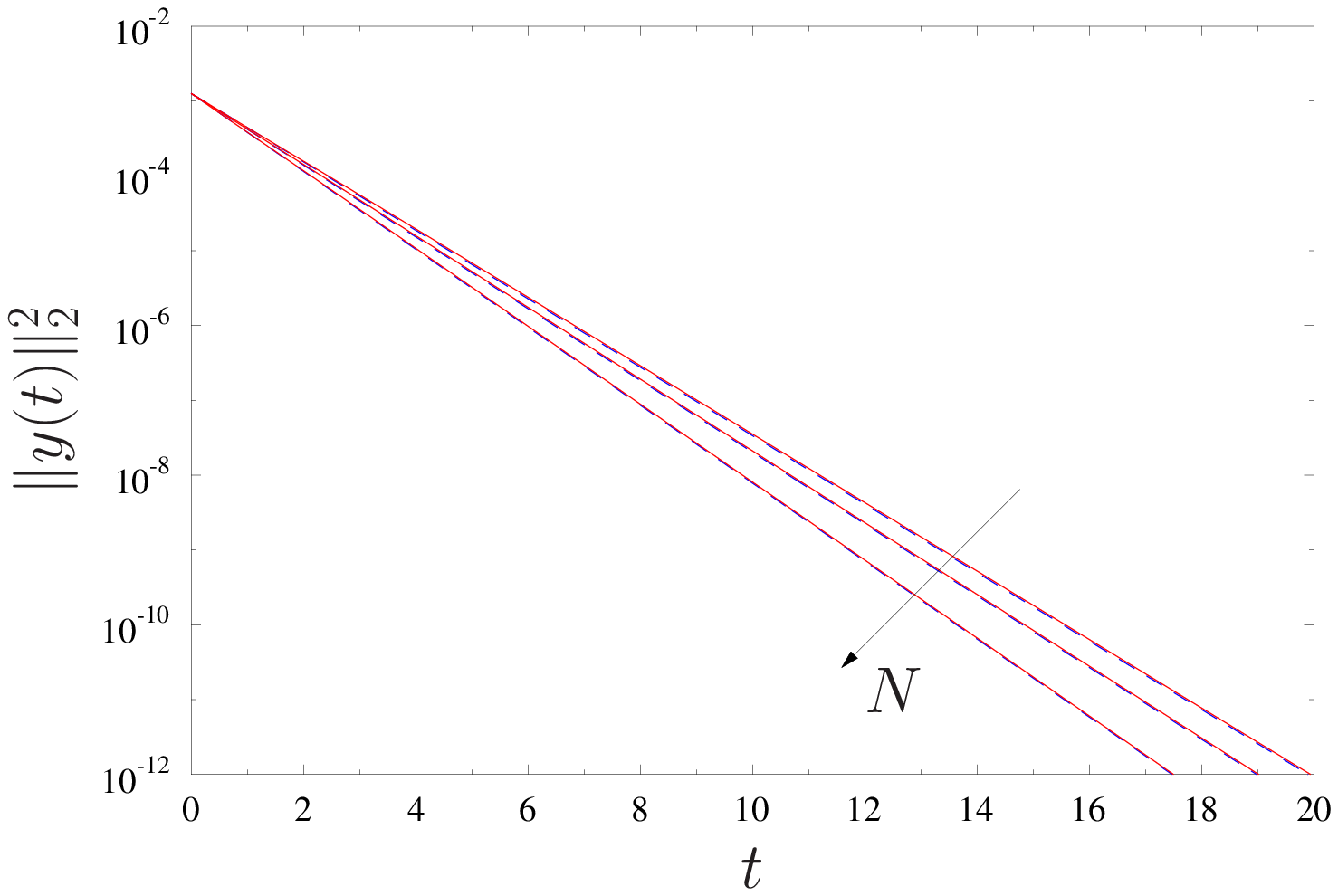}}
\quad
\subfigure[]{\includegraphics[width=0.5\textwidth]{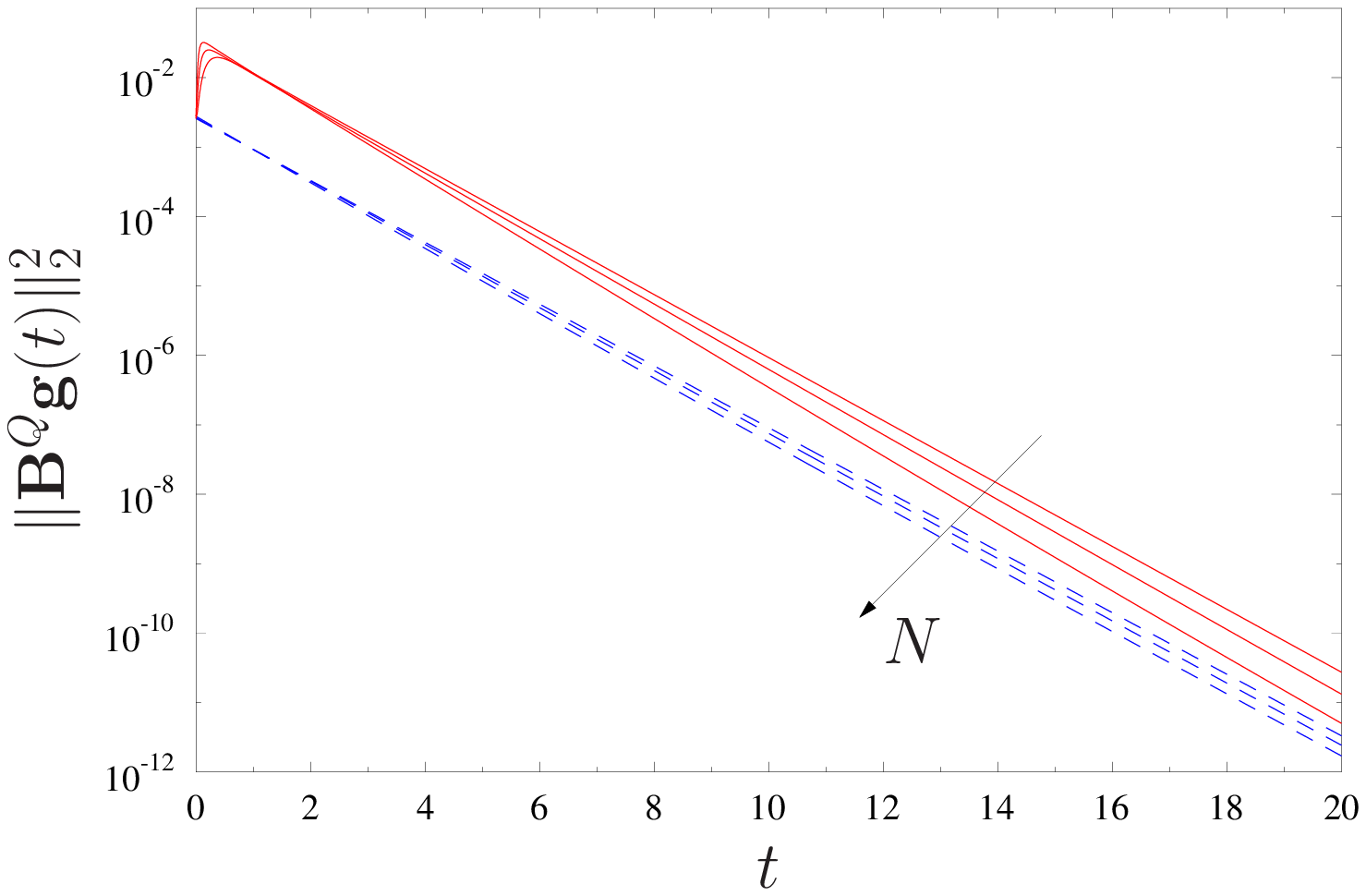}}}
\caption{Dependence of (a) the perturbation energy $\| y(t) \|_2^2$
  and (b) the \revt{actuation energy} $\| \B^{Q} \g(t)\|_2^2$ on time
  $t$ for a system discretized with increasing resolutions \revt{$N =
    15, 31, 63$} (the trends with increasing $N$ are indicated with
  arrows).  The initial perturbation, cf.~\eqref{eq:zeta0}, is scaled
  such that \revt{$\| \varepsilon\zeta_0 \|_{\infty} = 0.02$}, whereas
  the \revt{actuation uses the smallest possible number of sink/source
    pairs $N_c = N+1$, cf.~\eqref{eq:nNc}}. The data for the linear
  and nonlinear problems are shown with dashed blue and solid red
  lines, respectively.}
\label{fig:control_Nx}
\end{figure}

Next we analyze the ``continuous limit'' of the problem when the
resolution $N$ and the number $N_c$ of the \revt{sink/source} pairs
both increase such that \revt{$N_c = N+1$}. We note that in the limit
$N_c \rightarrow \infty$ each of the \revt{actuator arrays} may be
regarded as approximating \revt{a continuous distribution of
  blowing-and-suction on hypothetical walls, cf~\S\ref{sec:actuation}
  and figure \ref{fig:setting}.} The history of the perturbation
energy $\| y(t) \|_2^2$ and of the corresponding \revt{actuation
  energy} $\| \B^{Q} \g(t) \|_2^2$, cf.~\eqref{eq:g}--\eqref{eq:J},
are shown for different values of $N$ and $N_c$ for the linear and
nonlinear case in figures \ref{fig:control_Nx}(a) and
\ref{fig:control_Nx}(b), respectively. In the first figure we see that
\revt{there is very little difference in the behavior between the
  linear and nonlinear case, and in both cases} the perturbation
energy $\| y(t) \|_2^2$ vanishes exponentially with the rate of decay
increasing with $N$ (and therefore also with $N_c$).  \revt{On the
  other hand, in figure \ref{fig:control_Nx}(b) we see that the
  required control effort in the nonlinear case is larger than in the
  linear case. In addition, while in the linear case the
  \revt{actuation energy} vanishes exponentially for all times $t>0$,
  in the nonlinear case it first increases before eventually decaying
  exponentially with rates increasing with $N_c = N+1$. We remark that
  the data presented in figures \ref{fig:control_Nx}(a) and
  \ref{fig:control_Nx}(b) corresponds to initial perturbations with
  magnitude $\|\varepsilon\zeta_0\|_{\infty} = 0.02$ which is close to
  the magnitude of the largest perturbations which can be stabilized
  in the nonlinear case using the smallest possible number of actuator
  pairs $N_c = N + 1$, cf.~\eqref{eq:nNc}.}

\begin{figure}
\centering
\mbox{
\subfigure[]{\includegraphics[width=0.5\textwidth]{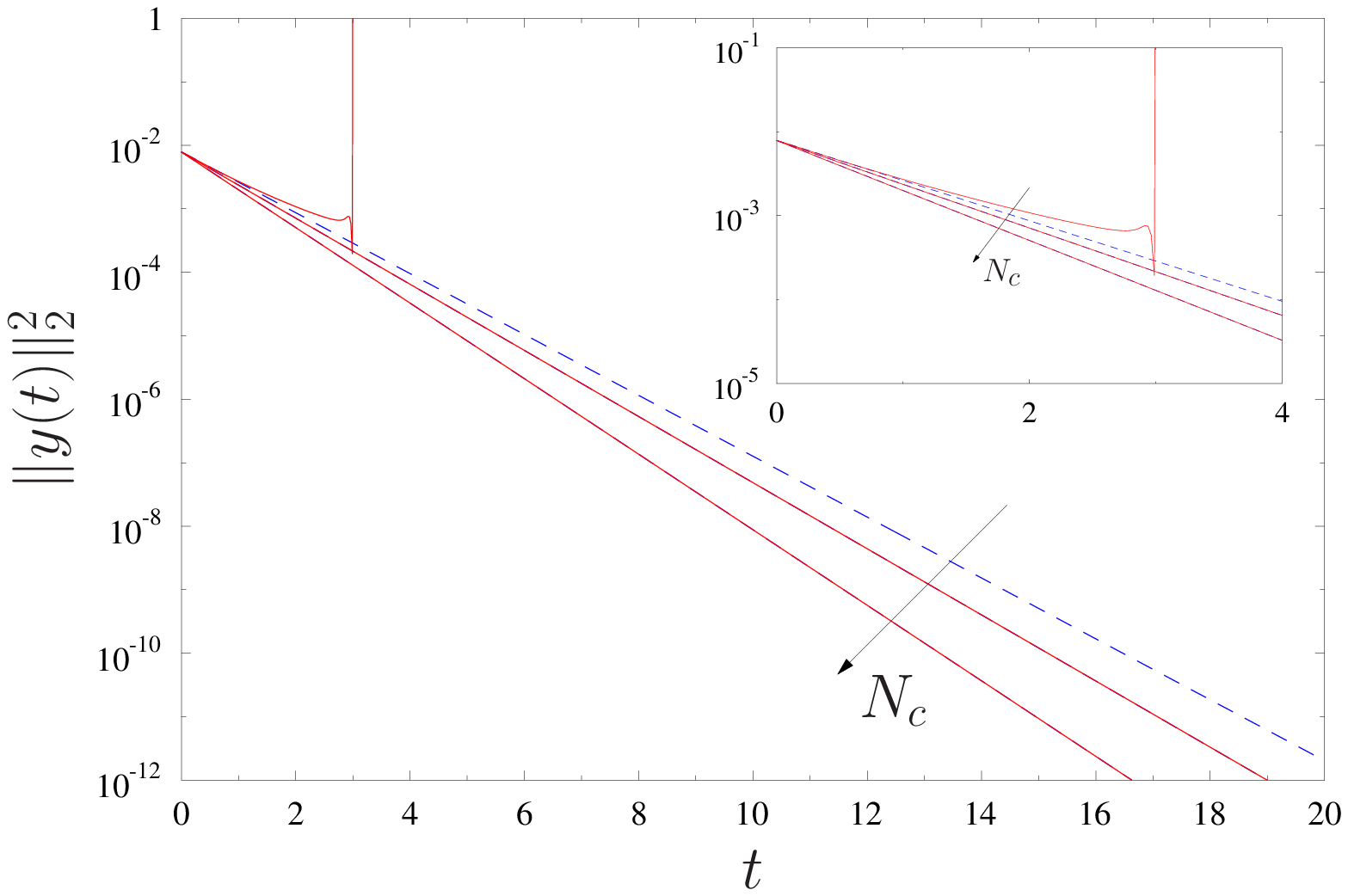}}
\quad
\subfigure[]{\includegraphics[width=0.5\textwidth]{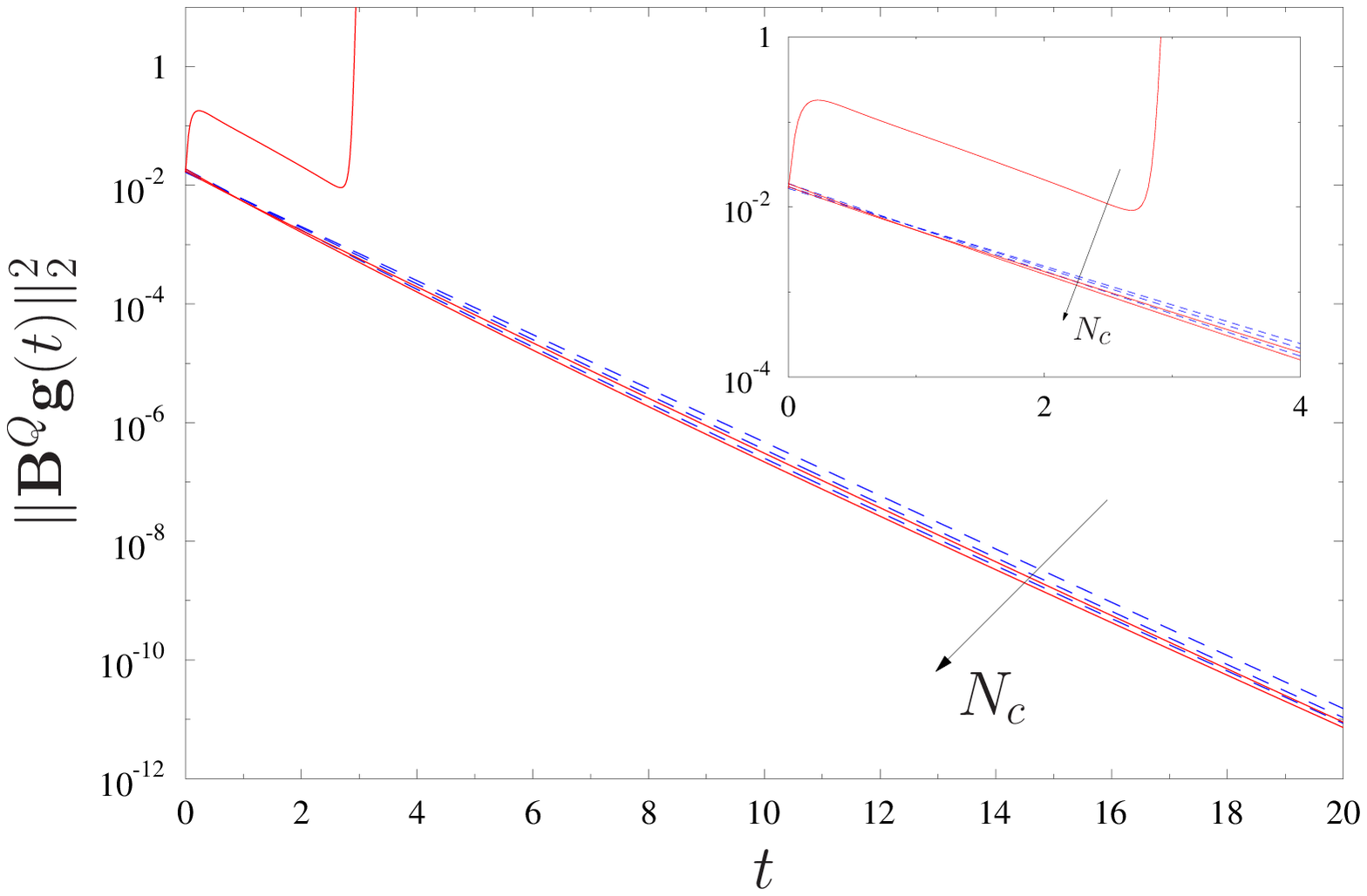}}}
\caption{Dependence of (a) the perturbation energy $\| y(t) \|_2^2$
  and (b) the \revt{actuation energy} $\| \B^{Q} \g(t) \|_2^2$ on time
  $t$ for a system discretized with the resolution \revt{$N = 31$} and
  with different numbers of \revt{actuator pairs $N_c = 32, 64, 128$}
  (the trend with increasing $N_c$ is indicated with arrows).
  \revt{The initial perturbation, cf.~\eqref{eq:zeta0}, is scaled such
    that $\| \varepsilon\zeta_0 \|_{\infty} = 0.05$.} The insets
  represent magnifications of the initial stages. The data for the
  linear and nonlinear problems are shown with dashed blue and solid
  red lines, respectively.}
\label{fig:control_Nc} 
\end{figure}

Now we move on to characterize the behavior of the linear and
nonlinear closed-loop systems \eqref{eq:ABK_N} and \eqref{eq:BRc3} as
the number $N_c$ of \revt{actuator} pairs is varied while the
resolution $N$ remains fixed.  The histories of the perturbation
energy $\| y(t) \|_2^2$ and of the \revt{actuation energy} \revt{$\|
  \B^{Q} \g(t)\|_2^2$} are shown in figures \ref{fig:control_Nc}(a)
and \ref{fig:control_Nc}(b), respectively, with insets offering
magnifications of the initial stages. The magnitude of the initial
perturbation, cf.~\eqref{eq:zeta0}, is \revt{$\| \varepsilon\zeta_0
  \|_{\infty} = 0.05$} in all cases. \revt{We see that in the linear
  problems the decay of the perturbation energy is exponential in all
  cases with the rate of decay increasing for larger $N_c$ and the
  same is also true for the behavior of the \revt{actuation energy}.
  On the other hand, in the nonlinear regime we observe that the
  perturbation cannot be stabilized when the smallest possible number
  $N_c = 32$ of actuator pairs is used and the solution blows up in
  finite time in that case. We note that with this number of actuators
  it was possible to stabilize perturbations with a smaller magnitude,
  cf.~figure \ref{fig:control_Nx}. This demonstrates that the
  nonlinear problem may not be stabilized if the magnitude of the
  initial perturbation $\varepsilon\zeta_0$ exceeds a certain
  threshold, even if the corresponding linear problem is controllable.
  When larger numbers of actuators are used, the perturbation energy
  in the solutions of the nonlinear closed-loop problems behaves
  similarly to its behavior in the solutions of the linear closed-loop
  problems. However, close inspection of figure
  \ref{fig:control_Nc}(b) indicates that when a larger number of
  actuator pairs is used, $N_c \ge 2(N+1)$, the \revt{actuation
    energy} required to stabilize the nonlinear system is in fact
  slightly smaller than the \revt{actuation energy} required to
  stabilize the corresponding linear system. As is evident from figure
  \ref{fig:control_Nx}(b), the opposite occurs when $N_c = N+1$.}

\begin{figure}
\centering
\mbox{
\subfigure[]{\includegraphics[width=0.475\textwidth]{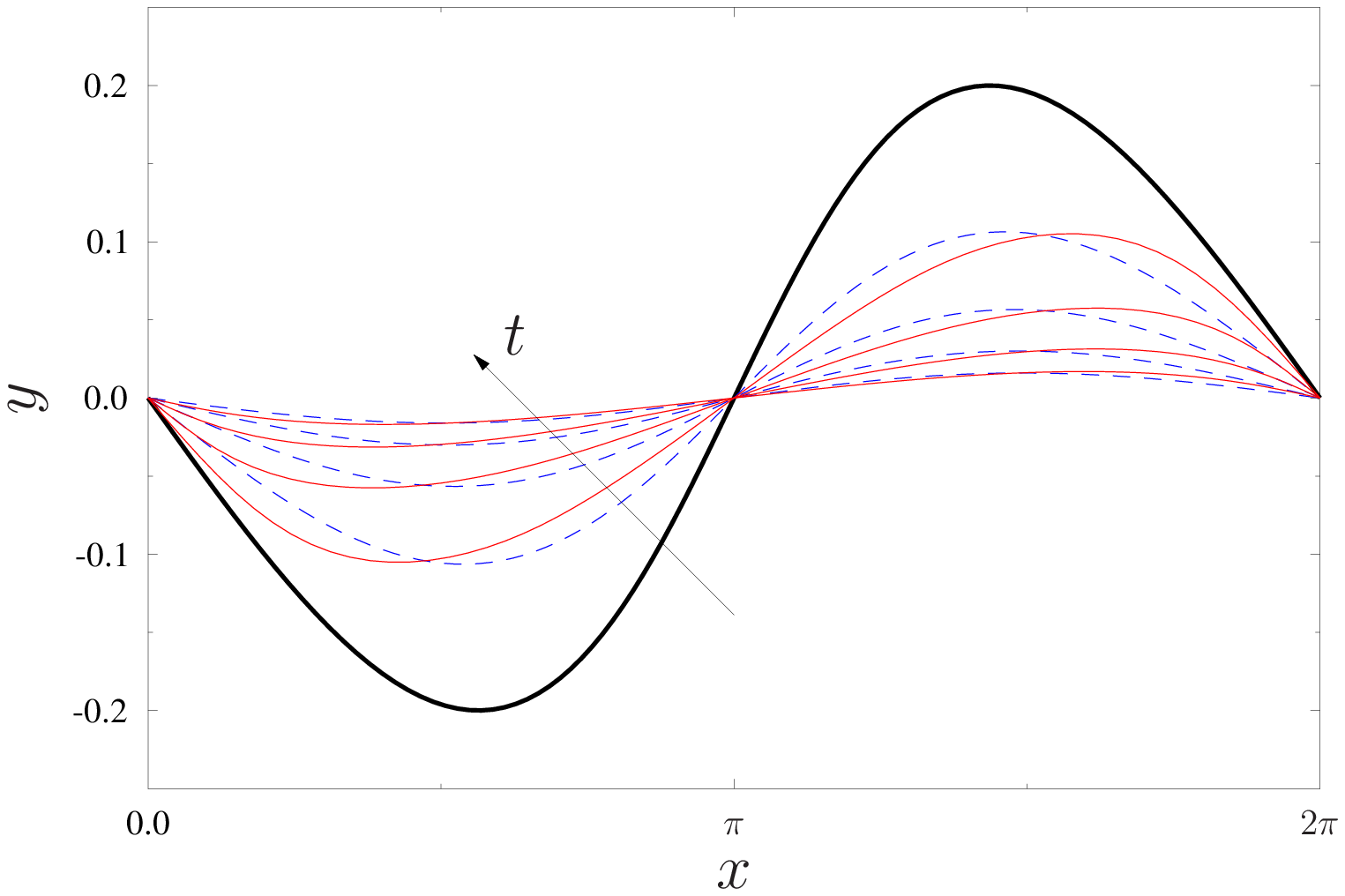}}
\quad
\subfigure[]{\includegraphics[width=0.475\textwidth]{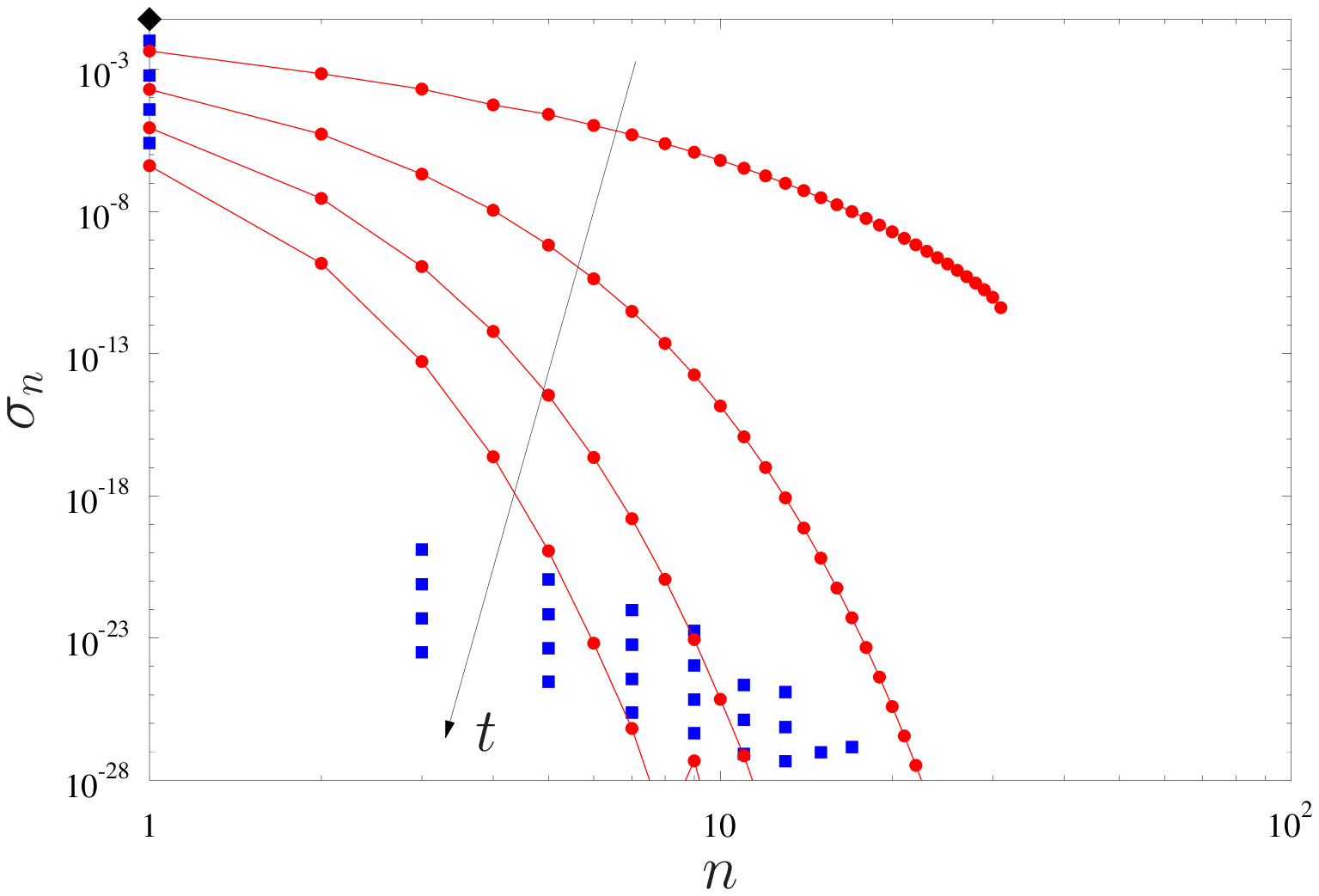}}}
\caption{Evolution of the solutions to the linear and nonlinear
  closed-loop problem \revt{obtained with resolution $N=31$,
    \revt{$N_c=64$} sink/source pairs and the initial data scaled such
    that \revt{$\|\varepsilon\zeta_0\|_{\infty} = 0.2$}} in (a) the
  physical space and (b) the Fourier space at a number of
  representative time instances \revt{(the trends with increasing $t$
    are indicated with arrows)}.  In panel (a) the data for the linear
  and nonlinear problem are shown with dashed blue and solid red
  lines, respectively, whereas the thick black solid line represents
  the initial condition $z_0$; in panel (b) {the data for the linear
    and nonlinear problem are shown with blue squares and and red
    circles, respectively, whereas the black diamond represents the
    initial condition $z_0$.}}
\label{fig:evolA}
\end{figure}

\begin{figure}
\centering
\mbox{
\subfigure[]{\includegraphics[width=0.45\textwidth]{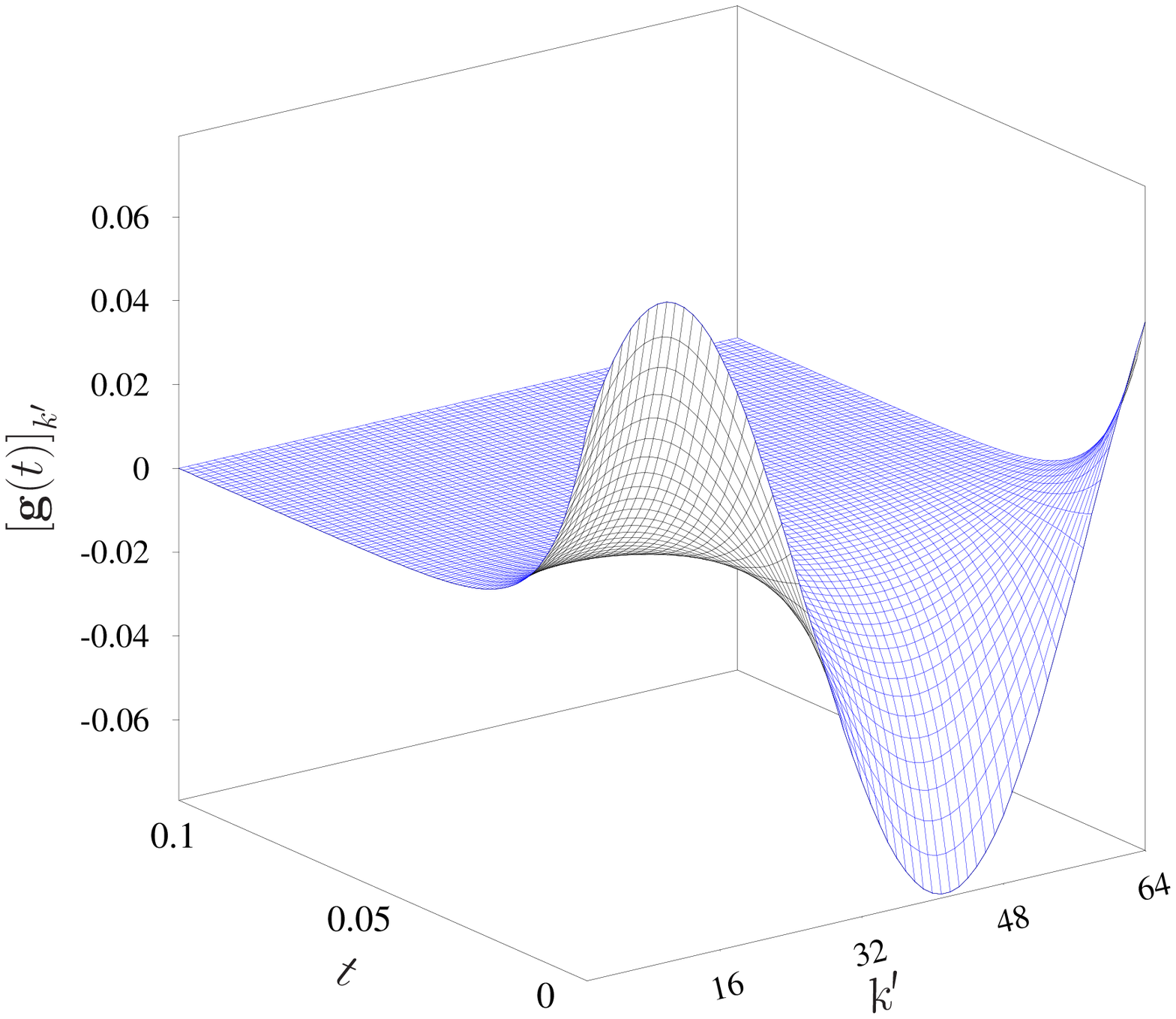}}
\quad
\subfigure[]{\includegraphics[width=0.45\textwidth]{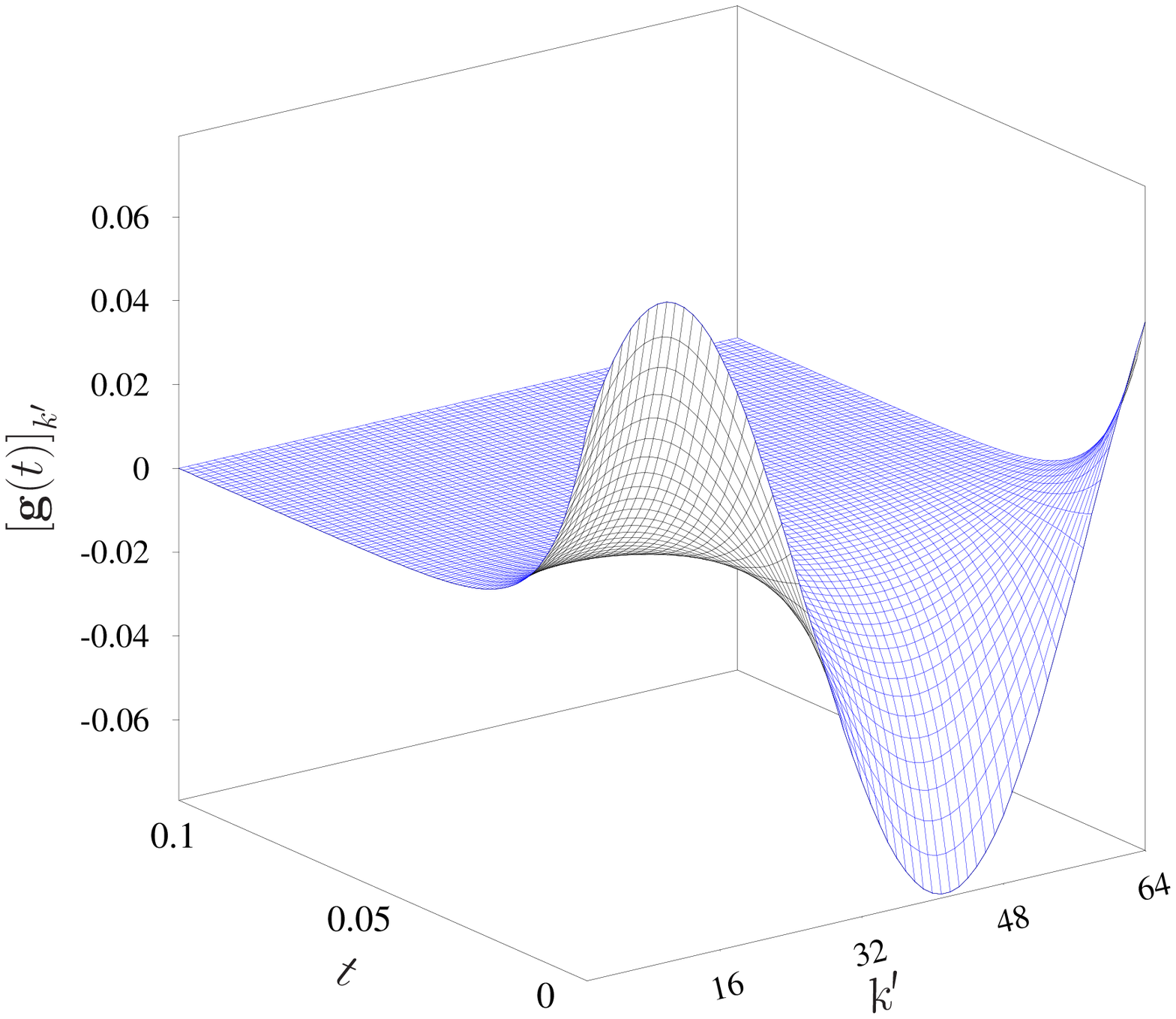}}}
\mbox{
\subfigure[]{\includegraphics[width=0.45\textwidth]{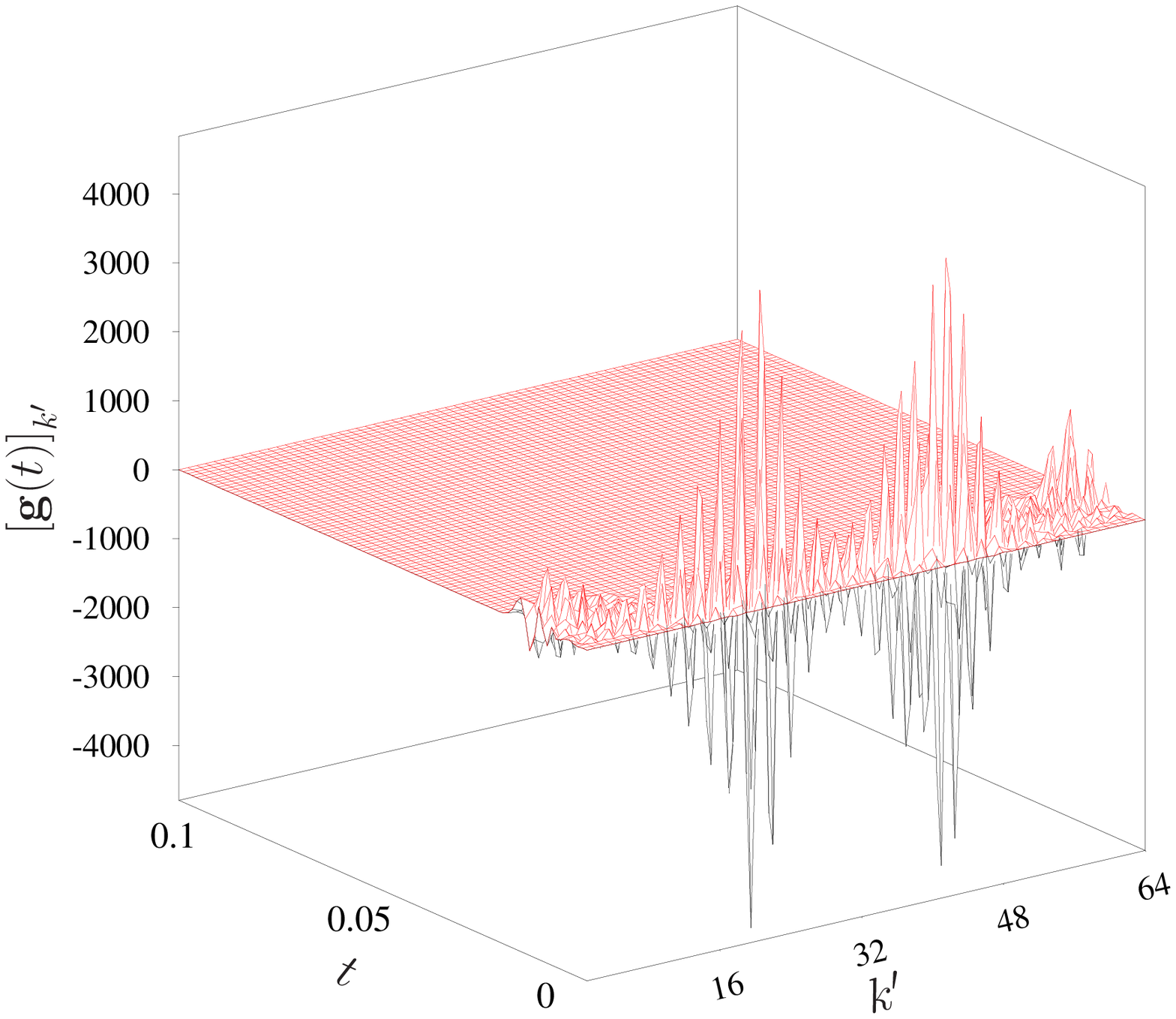}}
\quad
\subfigure[]{\includegraphics[width=0.45\textwidth]{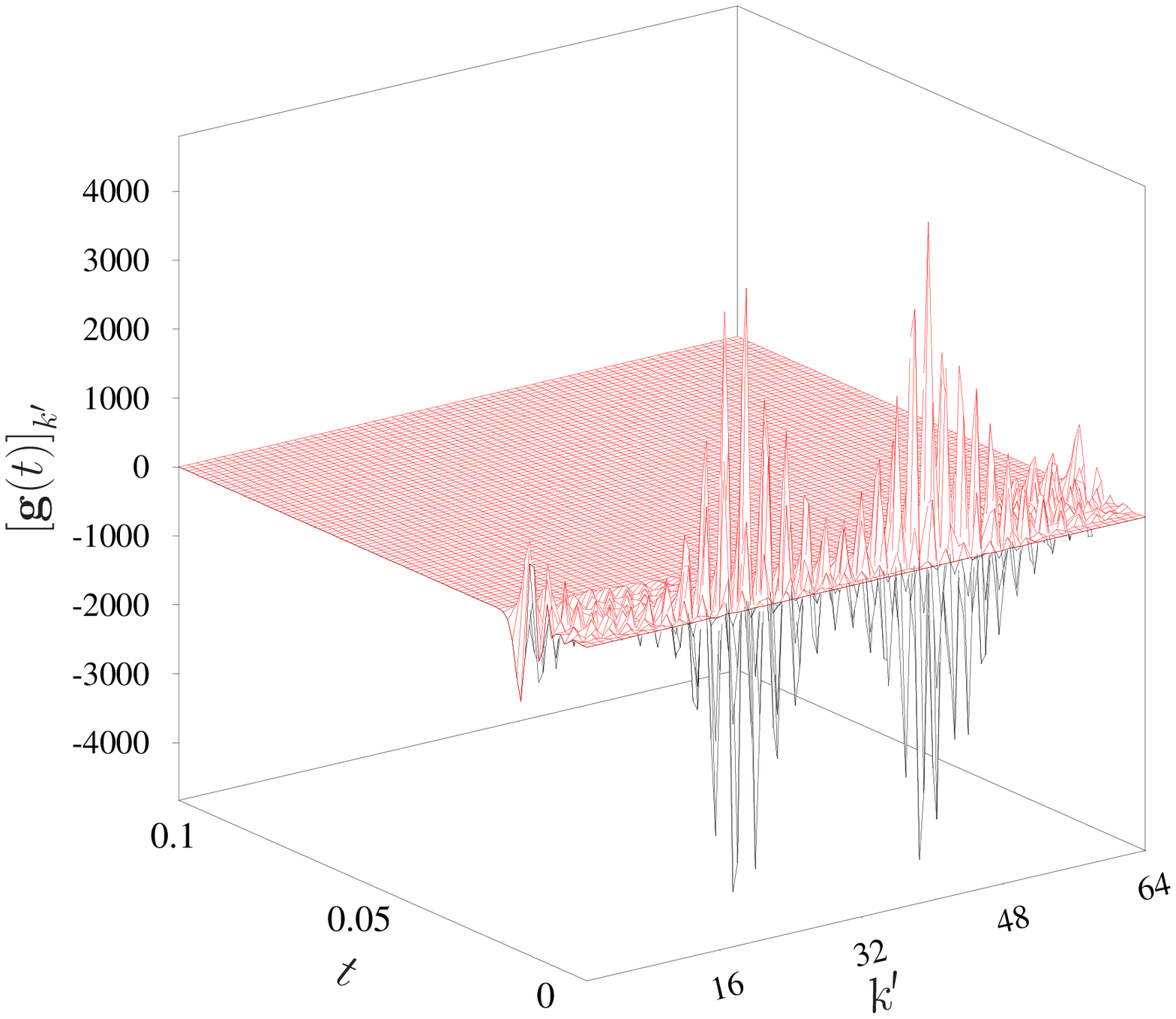}}}
\caption{\revt{Time histories of the intensities of individual
    actuators in the solution to (a,b) the linear and (c,d) the
    nonlinear closed-loop problem with resolution $N=31$ using
    \revt{$N_c = 64$} actuator pairs and the initial perturbation
    scaled such that \revt{$\|\varepsilon\zeta_0\|_{\infty} = 0.2$}.
    Panels (a,c) and (b,d) correspond to actuators in the upper and
    lower array (cf.~figure \ref{fig:setting}), where the actuator
    index is defined as $k' = (k+1) / 2$, $k=1,3,\dots,2N_c-1$, and
    $k' = k/2$, $k=2,4,\dots,2N_c$, respectively. For clarity, only
    short times \revt{$t \in [0,0.1]$} are shown.}}
\label{fig:Bu}
\end{figure}

\begin{figure}
\centering
\mbox{
\subfigure[]{\includegraphics[width=0.5\textwidth]{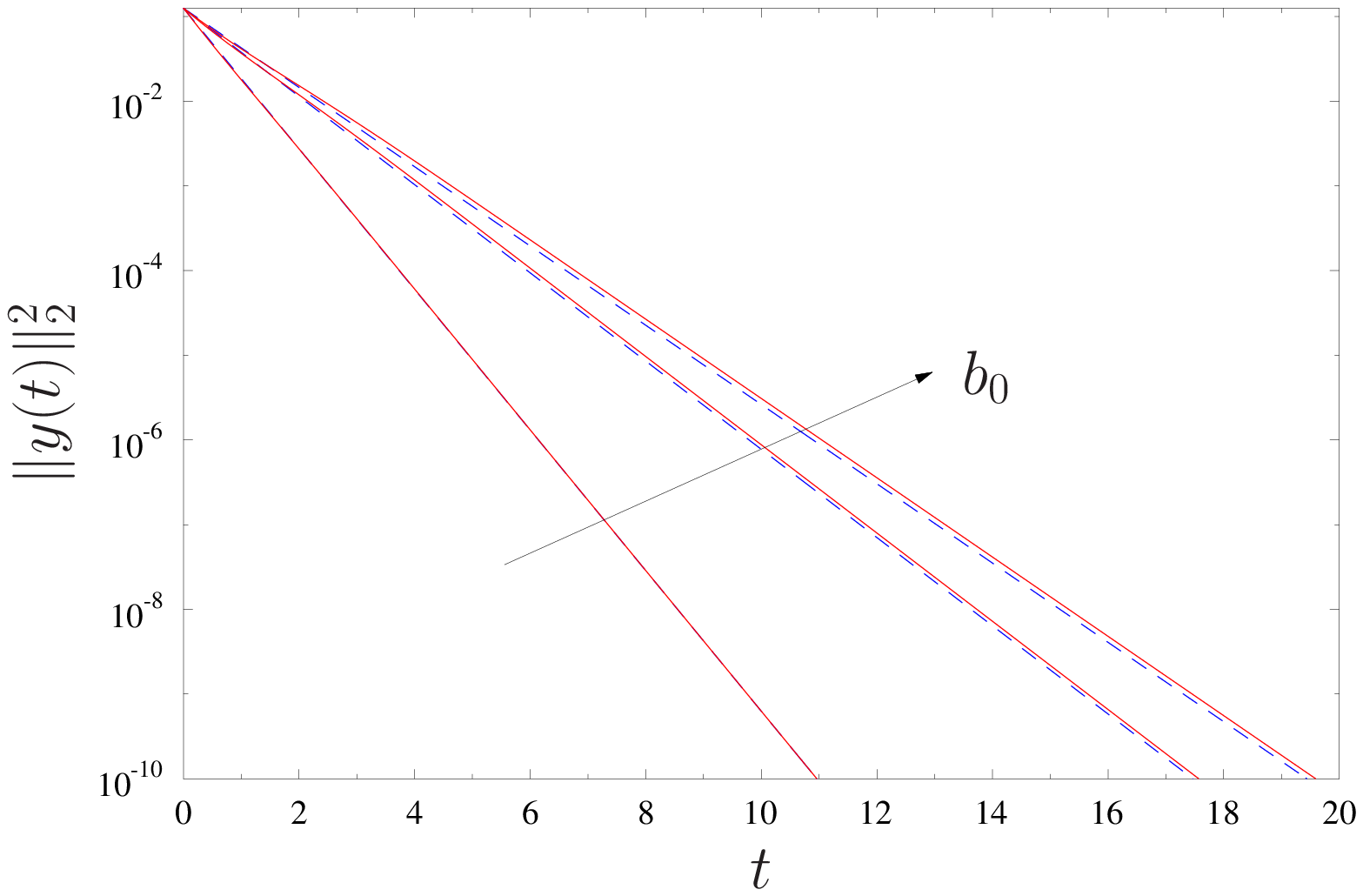}}
\quad
\subfigure[]{\includegraphics[width=0.5\textwidth]{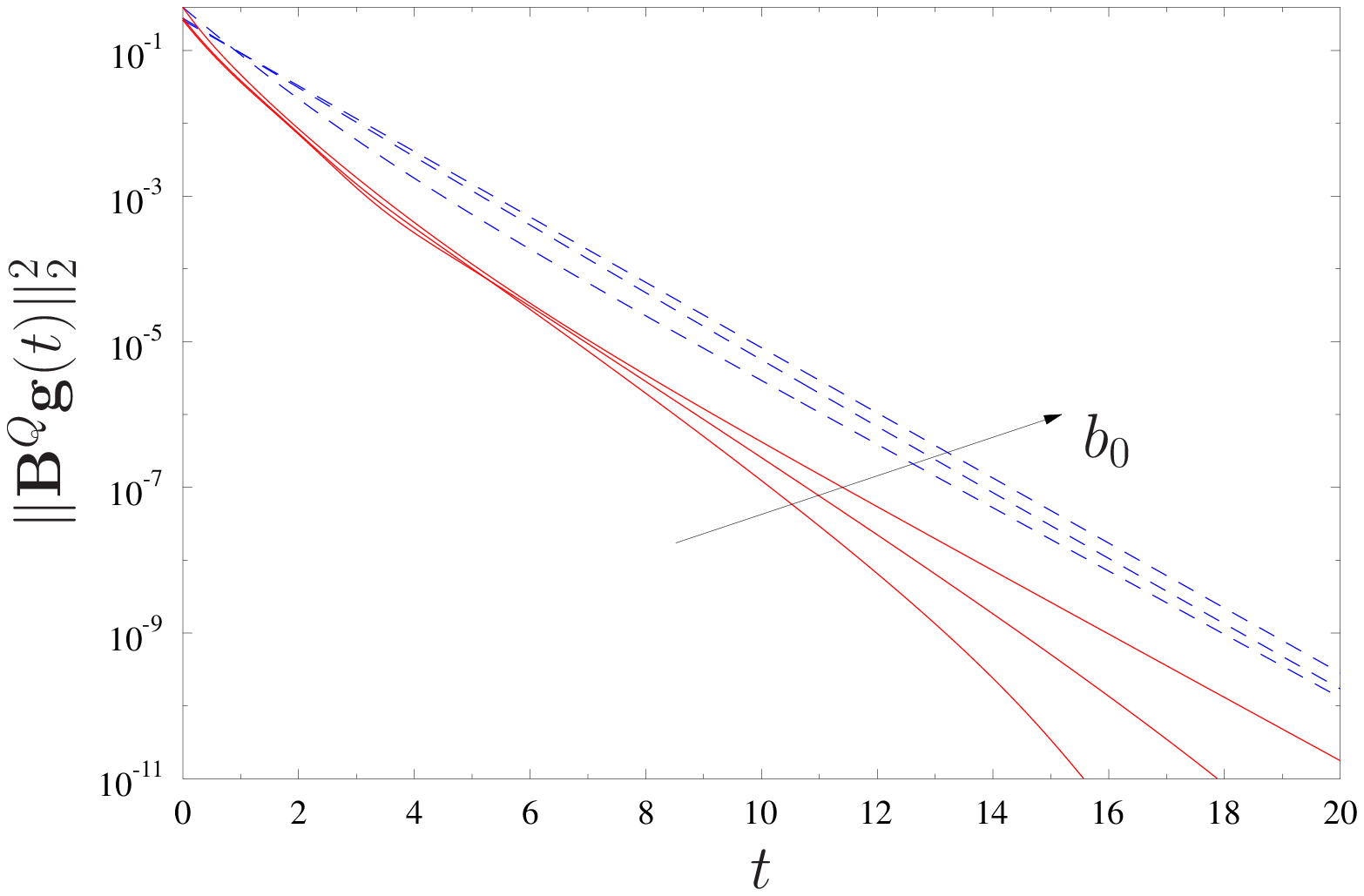}}}
\caption{Dependence of (a) the perturbation energy $\| y(t) \|_2^2$
  and (b) the \revt{actuation energy} $\| \B^{Q} \g(t)\|_2^2$ on time
  $t$ for different separations of the \revt{sinks/sources} from the
  sheet $b_0 = 0.1, 1, 1.5$ (the trends with increasing $b_0$ are
  indicated with arrows). The remaining parameters are \revt{$N=31$,
    $N_c=64$ and $\| \varepsilon\zeta_0 \|_{\infty} = 0.2$.}  The data
  for the linear and nonlinear problems are shown with dashed blue and
  solid red lines, respectively.}
\label{fig:control_b} 
\end{figure}

\begin{figure}
\centering
\mbox{
\subfigure[]{\includegraphics[width=0.5\textwidth]{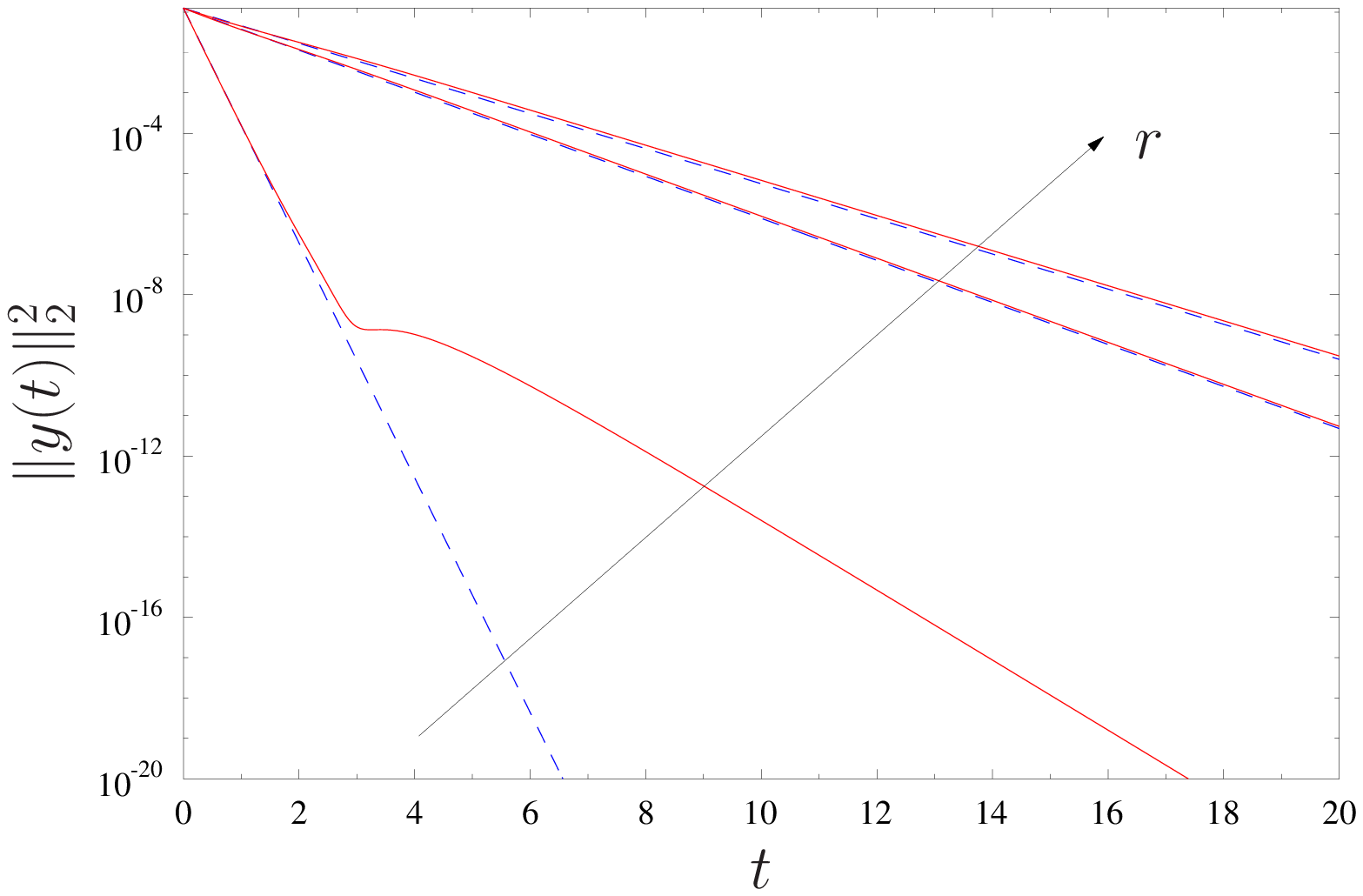}}
\quad
\subfigure[]{\includegraphics[width=0.5\textwidth]{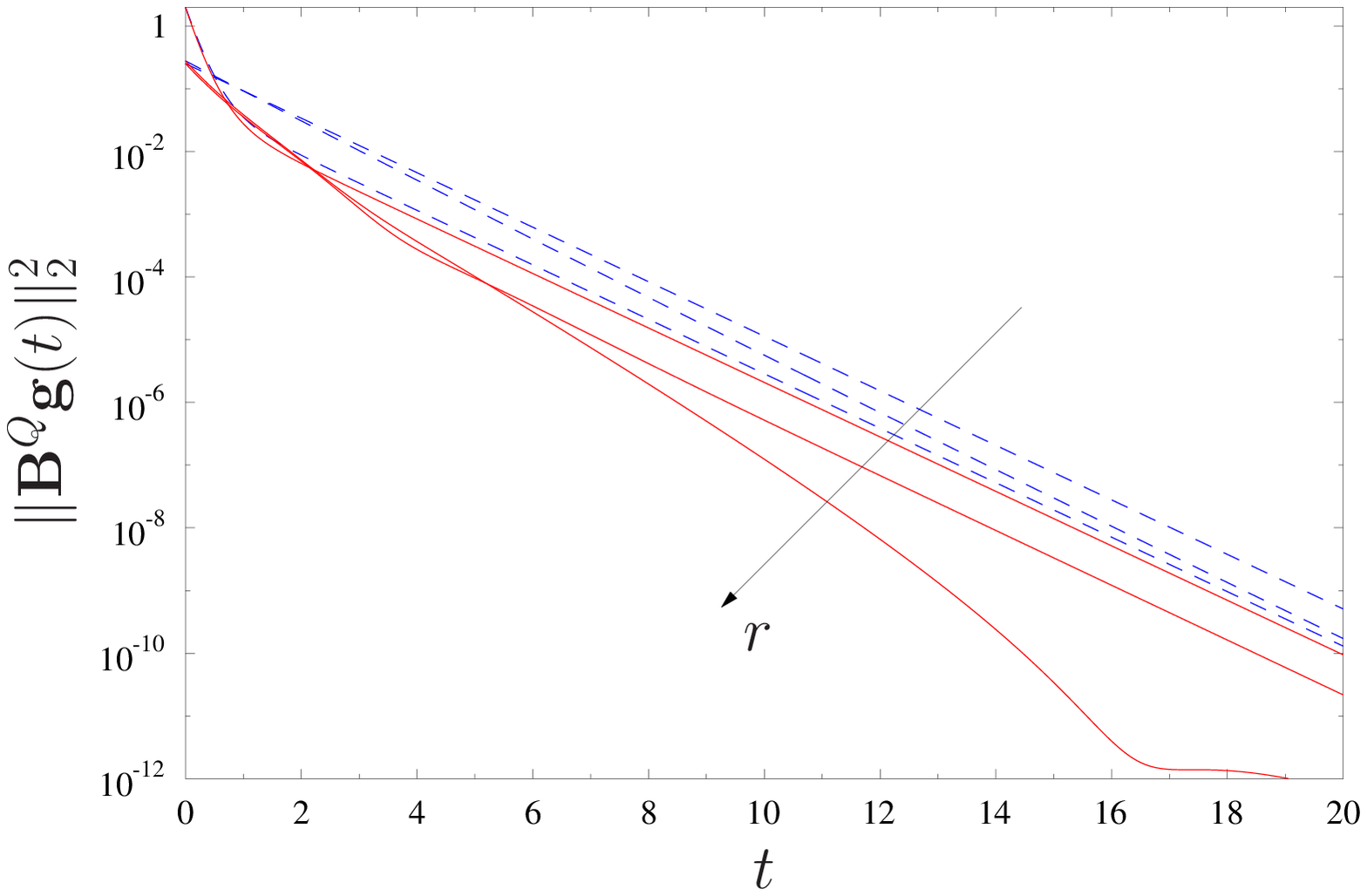}}}
\caption{Dependence of (a) the perturbation energy $\| y(t) \|_2^2$
  and (b) the \revt{actuation energy} $\| \B^{Q} \g(t)\|_2^2$ on time
  $t$ for different costs of control $r = 10^{-2}, 1, 10^2$ (the
  trends with increasing $r$ are indicated with arrows).  The
  remaining parameters are \revt{$N=31$, $N_c=64$ and $\|
    \varepsilon\zeta_0 \|_{\infty} = 0.2$.}  The data for the linear
  and nonlinear problems are shown with dashed blue and solid red
  lines, respectively.}
\label{fig:control_r} 
\end{figure}

\revt{To fix attention, in the remainder of this section we consider
  the case with the resolution $N=31$, $N_c = 64$ actuator pairs
  (which is twice as many as the minimum number required for
  controllability in the linear case for the given resolution $N$,
  cf.~\eqref{eq:nNc}) and the initial perturbation scaled such that
  $\| \varepsilon\zeta_0 \|_{\infty} = 0.2$. We now analyze the time
  evolution of the solution to the representative problem defined
  above} \revt{in the ``physical'' space in figure \ref{fig:evolA}(a)
  and in the Fourier-space representation in terms of the quantity
  $\sigma_n = (1/4)\left(
    |\alpha_{-n}|+|\beta_n|+|\beta_{-n}|+|\alpha_n|\right)$,
  $n=1,\ldots,N$, in figure \ref{fig:evolA}(b).  In figure
  \ref{fig:evolA}(a) we note that, as the control starts to act, the
  initial perturbation is attenuated without much change of shape,
  both in the linear and nonlinear setting. This behavior is also
  reflected in the corresponding evolution in the Fourier space
  {shown} in figure \ref{fig:evolA}(b), where we see that the control
  attenuates the spectrum of the solution uniformly across all
  wavenumbers and in the linear problem there is little energy
  {transfer} from the mode with $n_0 = 1$ to modes with higher
  wavenumbers. This latter property is a consequence of the fact that
  the closed-loop linear system is only weakly non-normal. In the
  nonlinear case the amplitude of the Fourier components with
  wavenumbers $n>1$ is larger than in the linear case, which is a
  consequence of the nonlinear energy transfer. Next, in figure
  \ref{fig:Bu} we present the corresponding histories of individual
  actuator intensities in the form of ``cascade plots''.} \revt{We see
  that in the linear problem the intensities of actuators in both the
  upper and lower array, respectively, $[g_1,g_3,\dots,g_{2N_c-1}]$
  and $[g_2,g_4,\dots,g_{2N_c}]$, at all times exhibit a smooth,
  wave-like dependence on the actuator index (which is proportional to
  its $x$-coordinate).  There is a certain phase shift between
  intensities of actuators in the two arrays. On the other hand, in
  the nonlinear problem the actuator intensities at early times reveal
  a fairly irregular dependence on the actuator index with neighboring
  actuators often having intensities of opposite signs, cf.~figures
  \ref{fig:Bu}(c,d).  However, the magnitudes of the actuator
  intensities in the nonlinear problem also exhibit a wave-like
  envelope similar to the behavior observed in the linear problem
  in~figures \ref{fig:Bu}(a,b). At initial times the actuator
  intensities in the nonlinear problem are significantly larger, by a
  few orders of magnitude, than the corresponding actuator intensities
  in the linear problem, although this difference vanishes as the
  magnitude $\| \varepsilon\zeta_0 \|_{\infty}$ of the initial
  perturbation decreases (this result is not shown here).
  Interestingly, the corresponding energy $\| \B^{Q} \g(t) \|_2^2$ of
  the velocity field induced by the actuation at the vortex sheet in
  its undeformed configuration is actually smaller in the nonlinear
  problem than in the linear case, see figures \ref{fig:control_b}(b)
  and \ref{fig:control_r}(b) discussed below.  The reason is that when
  neighboring actuators have intensities of opposite signs, this
  effect is canceled when the velocity they induce at the sheet
  location is computed as $\B^{Q} \g(t)$.}

In the remaining parts of this section we will study how the
performance of the closed-loop control in the linear and nonlinear
setting is influenced by a number of parameters characterizing the
problem.  First, we investigate the effect of the parameter $b_0$
describing the separation of the actuators from the vortex sheet in
its unperturbed configuration, cf.~\S \ref{sec:actuation}.  In figure
\ref{fig:control_b}(a) we see that, as expected, increasing the
separation $b_0$ results in a reduction of the exponential decay rate
of the perturbation energy in the linear closed-loop system, and
\revt{an essentially identical} trend is also evident in the evolution
governed by the nonlinear closed-loop system.  In figure
\ref{fig:control_b}(b) we observe that increasing the separation $b_0$
also tends to increase the \revt{actuation energy} $\| \B^{Q} \g(t)
\|_2^2$ required to stabilize the system \revt{with that energy again
  being smaller in the nonlinear case than in the linear case,
  cf.~figure \ref{fig:control_Nc}(b).}

The influence of the relative cost of control $r$, cf.~\eqref{eq:J},
on the performance of the control is documented in figure
\ref{fig:control_r}.  In figure \ref{fig:control_r}(a) we see that, as
expected, increasing the cost $r$ results in a reduction of the
exponential decay rate of the perturbation energy in the linear
closed-loop system, and \revt{similar trends are} also evident in the
evolution governed by the nonlinear closed-loop system, \revt{except
  for the case with $r=10^{-2}$ when at later times the convergence
  slows down. This can be attributed to the fact that with a low cost
  of control the actuation may act more ``aggressively'' which may not
  be desirable in the nonlinear setting.}  In figure
\ref{fig:control_r}(b) we note that increasing the value of $r$ tends
to reduce the \revt{actuation energy} $\| \B^{Q} \g(t) \|_2^2$
required to stabilize the system \revt{with that energy again being
  smaller in the nonlinear case than in the linear case, cf.~figure
  \ref{fig:control_Nc}(b). We also note that} the variation of the
\revt{actuation energy} is not very significant, even though the
considered values of $r$ span four orders of magnitude, \revt{which may
  suggest that in actual applications it may be advisable to consider
  ways of penalizing the control effort other than proposed in
  \eqref{eq:J}--\eqref{eq:J2}.}

\begin{figure}
\centering
\mbox{
\subfigure[]{\includegraphics[width=0.5\textwidth]{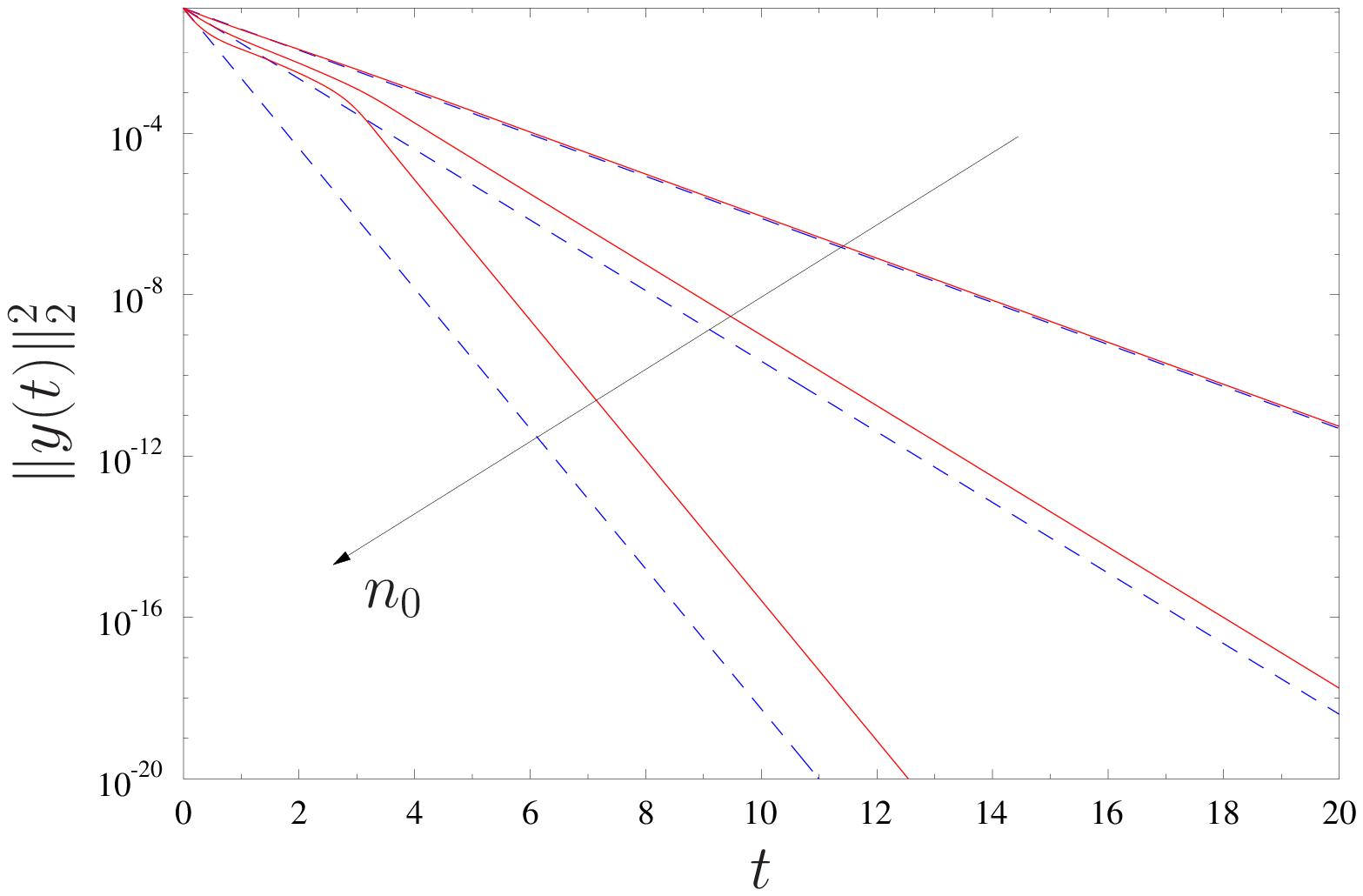}}
\quad
\subfigure[]{\includegraphics[width=0.5\textwidth]{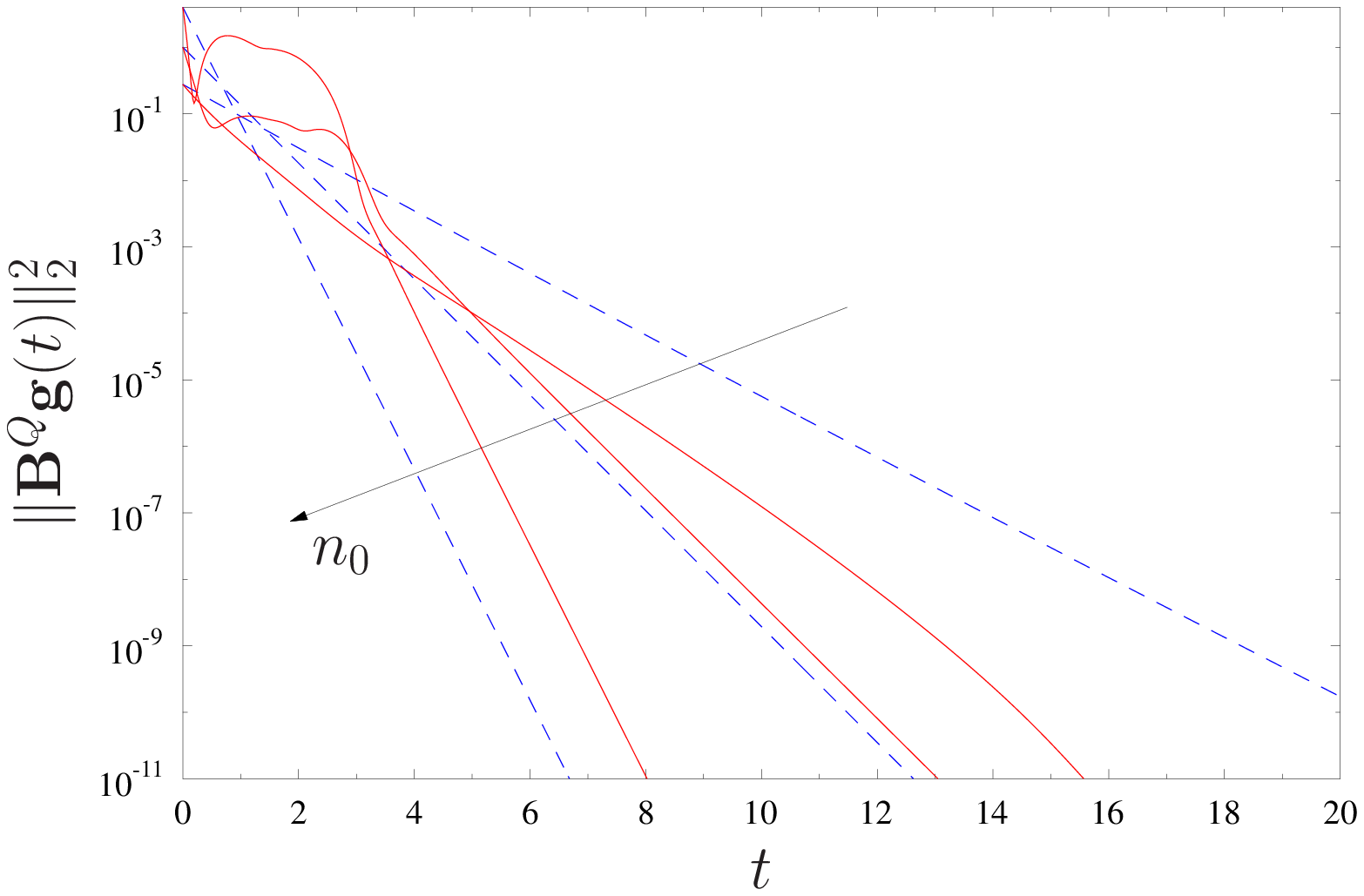}}}
\caption{Dependence of (a) the perturbation energy $\| y(t) \|_2^2$
  and (b) the \revt{actuation energy} $\| \B^{Q} \g(t)\|_2^2$ on time
  $t$ for initial perturbations $\zeta_0$ with different wavenumbers
  $n_0 = 1, 2, 4$ (the trends with increasing $n_0$ are indicated with
  arrows).  The remaining parameters are \revt{$N=31$, $N_c=64$ and
    $\| \varepsilon\zeta_0 \|_{\infty} = 0.2$.} The data for the
  linear and nonlinear problems are shown with dashed blue and solid
  red lines, respectively. }
\label{fig:control_Nic} 
\end{figure}

\begin{figure}
\centering
\mbox{
\subfigure[]{\includegraphics[width=0.5\textwidth]{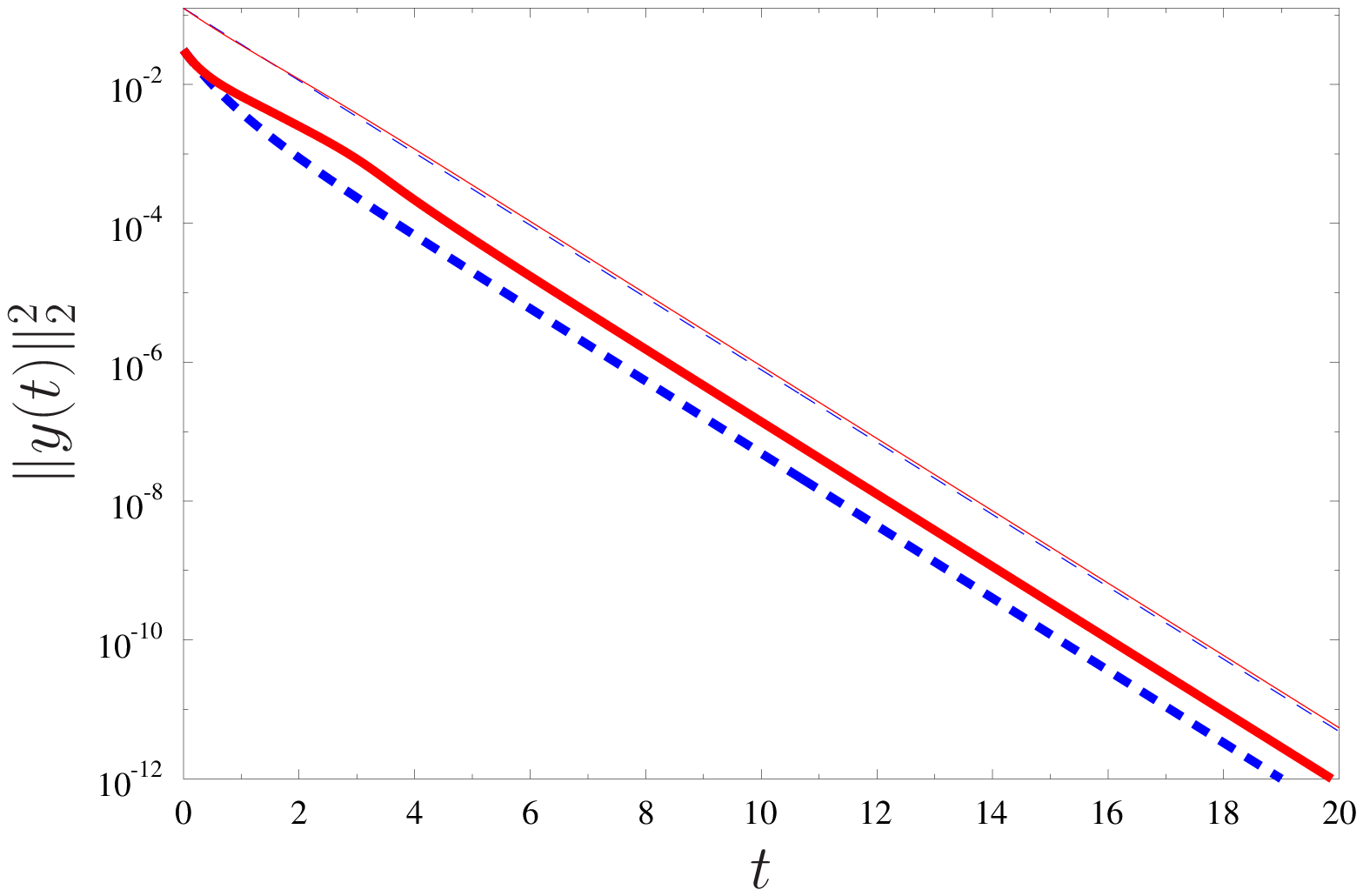}}
\quad
\subfigure[]{\includegraphics[width=0.5\textwidth]{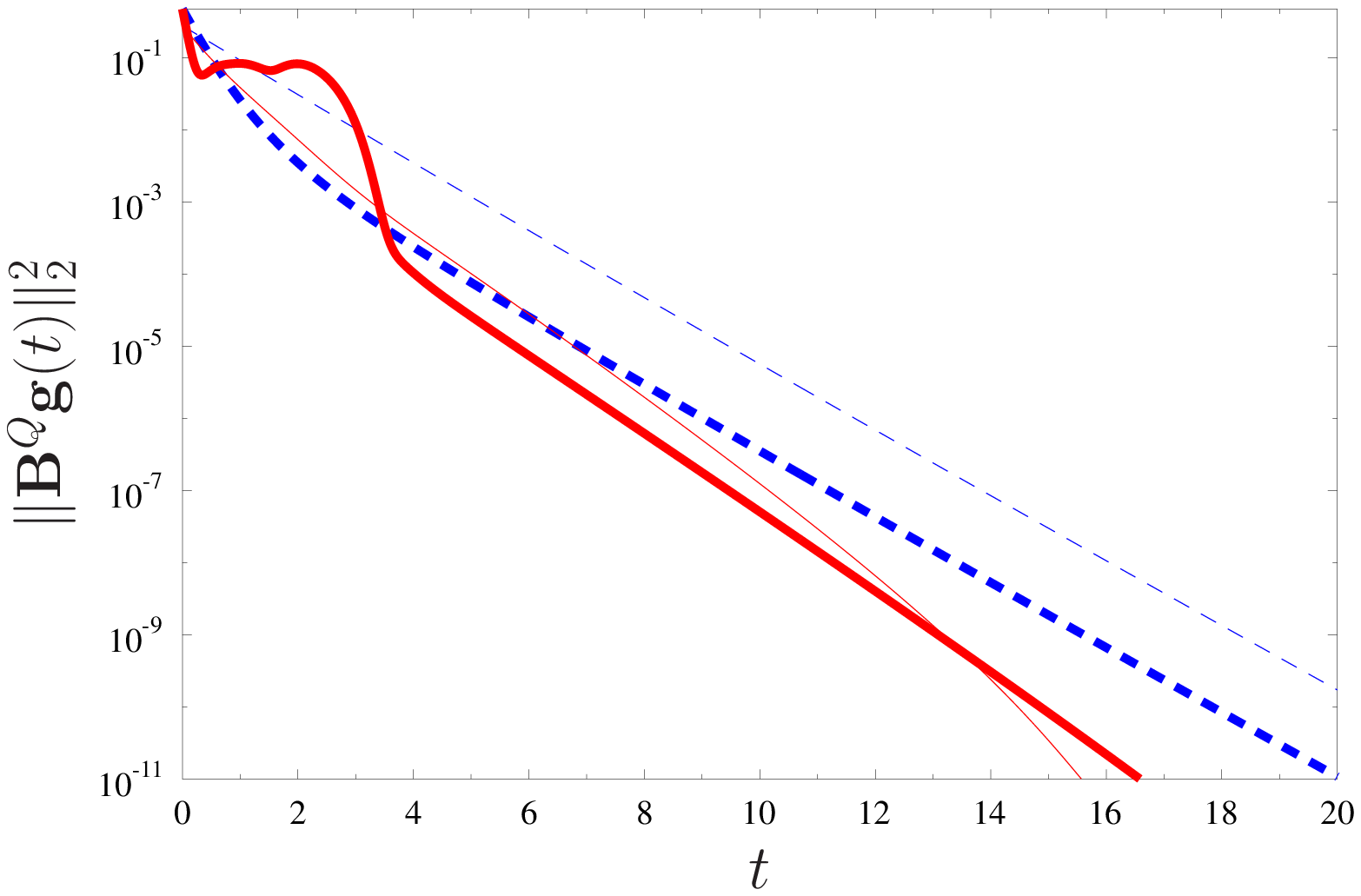}}}
\caption{\revt{Dependence of (a) the perturbation energy $\| y(t)
    \|_2^2$ and (b) the \revt{actuation energy} $\| \B^{Q}
    \g(t)\|_2^2$ on time $t$ for initial perturbations $\zeta_0$ with
    a single wavenumber $n_0=1$ (thin lines) and constructed as a
    superposition of components with wavenumbers $n_0 =1,2,3,4$ (thick
    lines).  The remaining parameters are \revt{$N=31$, $N_c=64$ and
      $\| \varepsilon\zeta_0 \|_{\infty} = 0.2$.} The data for the
    linear and nonlinear problems are shown with dashed blue and solid
    red lines, respectively. }}
\label{fig:control_nnNic} 
\end{figure}

We close this section by analyzing the performance of the linear and
nonlinear closed-loop systems \eqref{eq:ABK_N} and \eqref{eq:BRc3}
with initial perturbations characterized by different wavenumbers
$n_0$. \revt{We continue to focus on the case with resolution $N=31$
  and $N_c=64$ actuator pairs, but will now consider initial
  perturbations with wavenumbers $n_0=1,2,4$, all scaled such that $\|
  \varepsilon\zeta_0 \|_{\infty} = 0.2$.}  In figure
\ref{fig:control_Nic}(a) we see that the linear \revt{closed-loop}
evolution for the initial perturbations with \revt{different
  wavenumbers $n_0$} exhibits a purely exponential decay with rates
consistent with the dependence of the closed-loop eigenvalues on $n$
shown in figure \ref{fig:evals}.  \revt{In the corresponding nonlinear
  problems the decay of the perturbation energy is slightly slower,
  albeit still exponential, with the difference between the two cases
  increasing with $n_0$.  Analogous trends are also observed in figure
  \ref{fig:control_Nic}(b) as regards the dependence of the
  \revt{actuation energy} on time $t$ for initial perturbations with
  different wavenumbers $n_0$. Finally, in figure
  \ref{fig:control_nnNic} we analyze the performance of feedback
  stabilization in situations when the initial perturbation contains
  components with different wavenumbers, as happens in practice
  (cf.~figure \ref{fig:xi}). We consider the case when $\zeta_0$ is
  constructed as a superposition of Fourier components with
  wavenumbers $n_0 =1,2,3,4$ with equal magnitude and scaled such that
  $\| \varepsilon\zeta_0 \|_{\infty} = 0.2$, which is compared with
  the case when, as before, the initial perturbation has only one
  Fourier component with $n_0=1$. In figure \ref{fig:control_nnNic}(a)
  we see that in both cases the perturbation energy vanishes
  exponentially with the same rate which is determined by the
  slowest-decaying component (here, with $n_0=1$ present in both
  cases), cf.~figure \ref{fig:evals}. In the case with the
  multicomponent initial perturbation, the decay of the perturbation
  energy in the nonlinear problem is a bit slower than in the linear
  problem and is \revt{preceded} by an increase of the \revt{actuation
    energy} at the initial stages, cf.~figure
  \ref{fig:control_nnNic}(b).}

\FloatBarrier

\section{Discussion and Conclusions}
\label{sec:final}

In this study we have considered the problem of feedback stabilization
of a flat inviscid vortex sheet described by the Birkhoff-Rott
equation \eqref{eq:BR} as an idealized model for shear layers arising
in many important applications. In addition to serving as a prototype
of the Kelvin-Helmholtz instability which is central to many
vortex-dynamics phenomena, inviscid vortex sheets also give rise to
many interesting mathematical problems related to the ill-posedness of
the Birkhoff-Rott equation. While numerical solution of this equation
typically requires a suitable regularization, we focused here on an
admittedly harder problem when there is no regularization. There have
already been many successful applications of the methods of linear
control theory to flow problems \citep{Kim2007arfm,bhhs09}, but only
relatively few to inviscid vortex flows \citep{p08a,nps17}, as they
often give rise to a range of unique technical challenges, some of
which are reiterated below.

As the first main finding of our study, we demonstrated using
analytical computations that the Birkhoff-Rott equation linearized
around the flat-sheet configuration is controllable when \revt{the
  number $N_c$ of actuator pairs is sufficiently large relative to the
  number of degrees of freedom present in the truncated Fourier
  representation of the equation, more specifically, when condition
  \eqref{eq:nNc} is satisfied. This result holds when either point
  sinks/sources or point vortices are used as actuators and under the
  condition that their total mass flux or circulation is conserved in
  time independently for actuators located above and below the vortex
  sheet, cf.~\eqref{eq:const1}--\eqref{eq:const2}, which also
  guarantees global conservation of either mass or circulation. We
  focused here on actuation using point sinks/sources, because they
  represent typical experimental set-ups more accurately. However,
  feedback stabilization using point vortices as actuators gives
  essentially identical results. The controllability result holds for
  all generic actuator locations, except for the staggered arrangement
  shown in figure \ref{fig:actuators}(b) where a larger number of
  actuator pairs is needed to ensure controllability. Therefore, in
  our study we considered actuators in the aligned arrangement
  (cf.~figure \ref{fig:actuators}(a)).}  Based on \revt{this
  controllability} result, we designed a state-based LQR stabilization
strategy where the key difficulty was the numerical solution of the
Riccati equation \eqref{eq:Ric} needed to determine the feedback
kernel $\K$, which {is} extremely ill-conditioned even for modest
resolutions $N$.  This ill-conditioning is the result of the property
that the unstable modes have growth rates proportional to the
wavenumber $n$, while the authority the control actuation has over
these modes vanishes as $e^{-n}$, and was overcome by performing all
computations with increased arithmetic precision depending on the
resolution $N$. It should be emphasized here that, unlike some other
flow-control problems formulated in the spatially periodic setting
(e.g., \citet{Bewley1998jfm}), in the present case the problem of
determining the feedback kernels does not uncouple in the
\revt{Fourier} basis and the Riccati equation \eqref{eq:Ric} must be
solved for {\em all} wavenumbers $n=1,\ldots,N$ (the reason is that an
actuator with the general form \eqref{eq:C} simultaneously acts on
solution components with all wavenumbers). An alternative approach
would be therefore to consider actuation with two continuous
\revt{distributions of blowing-and-suction} whose intensities could be
represented in terms of a finite number of Fourier modes, however, the
formulation adopted in the present study is arguably more relevant and
at the same time more interesting.

Computations performed for the linear closed-loop system revealed
\revt{exponential decay of the perturbation energy and of the
  corresponding \revt{actuation energy} with the decay rates changing
  in the expected ways as different parameters are varied, namely, the
  number $N_c$ of actuator pairs, the distance $b_0$ between the
  actuators and the vortex sheet, the relative cost $r$ of the control
  and the wavenumber $n_0$ of the initial perturbation. In the
  nonlinear problems initial perturbations can be stabilized provided
  their magnitude is not too large, as otherwise a blow-up occurs in
  finite time (cf.~figure \ref{fig:control_Nc}(a)), although its
  mechanism appears quite different from the singularity formation
  occurring in the original Birkhoff-Rott equation \eqref{eq:BR}
  \citep{vsheet:Moore}.  More precisely, while in both cases it takes
  place at times $\O(1)$ for initial perturbations with the magnitude
  considered here, in the original Birkhoff-Rott equation
  \eqref{eq:BR} the singularity formation is signaled by the loss of
  analyticity by the solution, whereas in the problem with feedback
  stabilization \eqref{eq:BRc} the $L^2$ norm of the solution blows
  up. However, it is important to note that as the number $N_c$ of
  actuator pairs increases, initial perturbations with larger
  magnitudes can be stabilized. More specifically, initial
  perturbations with the transverse amplitude equal to more than 3\%
  of the sheet length could be effectively stabilized. This insight is
  the second main finding of this study.  The main objective of this
  investigation was to use a simple inviscid model of the vortex sheet
  dynamics given by the Birkhoff-Rott equation \eqref{eq:BR} to
  provide fundamental insights about the potential for stabilization
  of actual shear layers using feedback. Since with its ill-posedness
  the Birkhoff-Rott equation is a rather extreme example of
  instability, our findings indicate that, in principle, it may be
  possible to stabilize realistic shear layers, provided the actuation
  has sufficiently many degrees of freedom.}

\revt{When an initial perturbation can be stabilized in the nonlinear
  regime, the perturbation energy decays exponentially and the rate of
  decay is usually essentially the same as in the linear regime,
  except for cases when the cost of control $r$ is very small
  (cf.~figure \ref{fig:control_r}(a)) and when the initial
  perturbations $\zeta_0$ have wavenumbers $n_0 > 1$ (cf.~figure
  \ref{fig:control_Nic}(a)). The corresponding \revt{actuation energy}
  is larger than in the linear regime when the smallest possible
  number $N_c = N + 1$ of actuator pairs is used (cf.~figure
  \ref{fig:control_Nx}(b)), but becomes smaller than that when the
  number of actuator pairs is increased (cf.~figure
  \ref{fig:control_Nc}(b)). It would be interesting to characterize
  the energetic efficiency of the proposed stabilization strategy in
  terms of the ratio of the kinetic energies of the flow perturbation
  and the actuation. However, due to limitations of our simple
  potential-flow model, the latter quantity is not well defined here,
  because the velocity field induced by a point singularity is not
  square-integrable over the flow domain, hence such a quantitative
  comparison is not possible. We therefore used the quantities $\|
  y(t) \|_2^2$ and $\| \B^{Q} \g(t) \|_2^2$ as substitutes to
  represent, respectively, the perturbation and actuation energy.
  Likewise, the instantaneously large values of the control input
  $[g_1(t),\dots,g_{2N_c}(t)]$ observed in the nonlinear case for
  large initial perturbations, cf.~figures \ref{fig:Bu}(c,d), may also
  be regarded as an artifact related to modeling actuation in terms of
  point singularities.}

The perturbation energy was defined as a measure of the transverse
deformation of the sheet. Our computations (not reported here)
performed with an alternative definition of the perturbation energy
based on the $L^2$ norm of the entire perturbation $\zeta(t)$, i.e.,
also involving its real (longitudinal) component, showed a
qualitatively similar performance of the control, except that the
actuation was less efficient as it also had to attenuate the
longitudinal part of the perturbation (which is physically irrelevant
as it amounts to merely reparameterizing the equilibrium configuration
$\tilde{z}(\gamma)$).  Our analysis also demonstrated that the
proposed feedback stabilization strategy exhibits the expected
behavior as different parameters ($b_0$, $r$ and $n_0$) are varied,
\revt{although the dependence of the required \revt{actuation energy}
  on its relative cost $r$ is rather weak, indicating that in actual
  applications different ways of penalizing the control effort may be
  preferable.}

\revt{Let us finally discuss some future extensions of the present
  work towards more practical stabilization problems involving shear
  layers. The first step in this directions will be to investigate
  stabilization of vortex sheets under the influence of viscosity or
  surface tension~\citep{hls97} and of vortex layers with a finite
  thickness~\citep{bs90}. It is known that both these effects may
  significantly affect the stability of the original problem.  For
  example, viscosity tends to suppress the growth of perturbations
  with higher wavenumbers such that the Kelvin-Helmholtz instability
  resulting from such perturbations is only moderate.  \revt{For some
    common regularizations the high-wavenumber instability becomes
    weaker and may be entirely eliminated, which means that unstable
    modes may be restricted to a certain finite range of wavenumbers,
    say $[N_1,N_2]$.  Then, if we set the number of actuator pairs as
    $N_c =2(N_2+1)$, the viscous vortex sheet should be stabilizable
    with less actuation.} As mathematical models of such effects we
  can utilize the dissipative $\delta$-regularization~\citep{k86a} or
  the dispersive $\alpha$-regularization~\citep{hnp06a} of vortex
  sheets which both reduce or suppress entirely the instability of
  higher-wavenumber modes (cf.~figure~3 in \cite{hnp06a}). In
  addition, since} state-based feedback stabilization strategies are
not very useful in practical situations where typically only
incomplete and often noisy measurements are available {instead of
  full-state information}, a more applicable approach would involve an
output-based compensator combining a state estimator, such as some
form of the Kalman filter, with LQR stabilization~\citep{Protas2004pf,
  nps17}. This constitutes a natural extension of the
\revt{state-feedback stabilization approach presented here} and, as a
step in this direction, in Appendix \ref{sec:observe} we address the
companion question concerning observability of the linearized
Birkhoff-Rott equation with pointwise velocity measurements.
\revt{Finally, a feedback stabilization (or compensation) approach
  designed based on such an extended model can be applied to stabilize
  the Kelvin-Helmholtz instability of viscous shear layers modeled by
  the Navier-Stokes equation \revt{where high-wavenumber perturbations
    are strongly suppressed by viscous effects}, in the spirit of what
  was done by \cite{Protas2004pf} in the context of the B{\'e}nard-von
  K{\'a}rm{\'a}n instability in the cylinder wake.}  Another related
problem concerns stabilization of inviscid vortex sheets in other
geometries such as the sphere \citep{vsheet:Sa04b}, which may have
interesting geophysical implications. {We add that, in some
  applications, such as for example mixing enhancement, one may be
  interested in stabilizing the system selectively, by damping
  perturbations only with certain wavenumbers; this represents yet
  another possible extension of this study.}

\section*{Acknowledgments}

\revt{An anonymous referee is acknowledged for providing important
  insights about the role of constraints
  \eqref{eq:const1}--\eqref{eq:const2}.}  The authors are grateful to
Pavel Holoborodko from Advanpix LLC for his assistance with
arbitrary-precision computations required to determine the feedback
operators. They also acknowledge the generous hospitality of the
Fields Institute for Mathematical Sciences in Toronto where this work
began during the Thematic Program on Multiscale Scientific Computing
(January--April, 2016). The first author acknowledges the JSPS
Fellowship which he held at Kyoto University in 2017 and partial
support through an NSERC (Canada) Discovery Grant. The second author
was supported by JSPS Kakenhi(B) \#26287023 and JSPS A3 Foresight
Program.

\appendix
\section{Spectral Representations of Kernels}
\label{sec:spectral}

Derivation of the spectral representation \eqref{eq:dzetadt} of the
linearized Birkhoff-Rott equation with control relies on the
Fourier-series expansion of the function $\cot(z)$ and its derivative.
More specifically, the Fourier coefficients $\widehat{\Phi}_n$ and
$\widehat{\Psi}_n$ of $\cot\left((x-a-ib)/2\right)$ and
$1/\sin^2((x-a-ib)/2)$ are defined through the following identities
\begin{align}
\cot\left(\frac{x-a-ib}{2}\right) &= 
\sum_{n=-\infty}^\infty \widehat{\Phi}_n \, \mbox{e}^{inx}, \label{eq:f1} \\
-\frac{1}{2\sin^2\left(\frac{x-a-ib}{2}\right)} &= 
\sum_{n=-\infty}^\infty \widehat{\Psi}_n \, \mbox{e}^{inx}. \label{eq:f2}
\end{align}
We will assume that $x, a\in \RR$ and will consider two cases
depending on whether $b \in \RR \backslash \{0\}$ is positive or
negative (we recall that $b$ represents the vertical location of an
actuator, cf.~\S \ref{sec:actuation}).

When $b > 0$, we have
\begin{align*}
\cot\left(\frac{x-a-i b}{2}\right) &= i \frac{\mbox{e}^{i(x-a-ib)/2} + \mbox{e}^{-i(x-a-ib)/2}}{\mbox{e}^{i(x-a-ib)/2} - \mbox{e}^{-i(x-a-ib)/2}} 
               = i \frac{1 + \mbox{e}^{-i(x-a-ib)}}{1 - \mbox{e}^{-i(x-a-ib)}} \\
               &= i\left( 1 + \frac{ 2 \mbox{e}^{-i(x-a-ib)} }{1 - \mbox{e}^{-i(x-a-ib)}} \right) 
               = i\left( 1 + 2 \mbox{e}^{-i(x-a-ib)} \frac{ 1 }{1 - \mbox{e}^{-i(x-a-ib)}} \right) \\
               &= i\left( 1 + 2 \mbox{e}^{-i(x-a-ib)} \sum_{n=0}^\infty \mbox{e}^{-in(x-a-ib)} \right) 
               =  i\left( 1 + 2 \sum_{n=1}^\infty \mbox{e}^{-in(x-a-ib)} \right) \\
               &= i\left( 1 + 2 \sum_{n=1}^\infty \mbox{e}^{-nb + ina} \mbox{e}^{-inx} \right).
\end{align*}
The infinite summation in the third equality is absolutely convergent
owing to the fact that $\left\vert \mbox{e}^{-i(x-a-bi)} \right\vert =
\mbox{e}^{-2b} <1$. Hence, the series is differentiable term by term, which
gives rise to the following spectral representation of $1/\sin^2((x-a-ib)/2)$
\begin{equation*}
-\frac{1}{2\sin^2\left( \frac{x-a-ib}{2} \right)} = i\left( 2 \sum_{n=1}^\infty (-in) \mbox{e}^{-nb + ina} \mbox{e}^{-inx} \right) 
 = 2 \sum_{n=1}^\infty n \mbox{e}^{-nb + ina} \mbox{e}^{-inx}.
\end{equation*}
On the other hand, when $b<0$, owing to the fact that $\left\vert
  \mbox{e}^{i(x-a-ib)} \right\vert = \left\vert \mbox{e}^b \right\vert
< 1$, we have
\begin{align*}
\cot\left(\frac{x-a-i b}{2}\right) &= i \frac{\mbox{e}^{i(x-a-ib)/2} + \mbox{e}^{-i(x-a-ib)/2}}{\mbox{e}^{i(x-a-ib)/2} - \mbox{e}^{-i(x-a-ib)/2}} 
               = i \frac{\mbox{e}^{i(x-a-ib)}+1}{\mbox{e}^{i(x-a-ib)}-1} \\
               &= i\left( 1 - \frac{2}{1 - \mbox{e}^{i(x-a-ib)}} \right) 
               = i\left( 1 - 2 \sum_{n=0}^\infty \mbox{e}^{in(x-a-ib)} \right) \\
               &= i\left( 1 - 2 \sum_{n=0}^\infty \mbox{e}^{nb-ina} \mbox{e}^{inx} \right) 
               = i\left( -1 - 2 \sum_{n=1}^\infty \mbox{e}^{nb-ina} \mbox{e}^{inx} \right) 
\end{align*}  
and 
\begin{equation*}
-\frac{1}{2\sin^2\left( \frac{x-a-ib}{2} \right)} = i\left( - 2 \sum_{n=1}^\infty (in) \mbox{e}^{nb-ina} \mbox{e}^{inx} \right) 
 = 2 \sum_{n=1}^\infty n \mbox{e}^{nb-ina} \mbox{e}^{inx}. 
\end{equation*}
In summary, the Fourier coefficients $\widehat{\Phi}_n$ and
$\widehat{\Psi}_n$, cf.~\eqref{eq:f1}--\eqref{eq:f2}, are for $b>0$
given by
\begin{eqnarray}
\widehat{\Phi}_n &=&
\left\{
\begin{array}{ll}
                   0, & n \ge 1,\\ 
                   i, & n=0, \\
2i\mbox{e}^{-\vert n \vert b+i\vert n \vert a},& n \le -1, \\
\end{array}
\right. \label{F-cot-p}\\
\widehat{\Psi}_n &=&
\left\{
\begin{array}{ll}
                   0, & n \ge 0,\\ 
2\vert n \vert \mbox{e}^{-\vert n \vert b+i\vert n \vert a},& n \le -1, \\
\end{array}
\right. \label{F-sin2-p}
\end{eqnarray}
and for $b<0$ {by}
\begin{eqnarray}
\widehat{\Phi}_n &=&
\left\{
\begin{array}{ll}
  -2i\mbox{e}^{ n b-i n a},& n \ge 1, \\
                   -i, & n=0, \\
                    0, & n \ge 1,
\end{array}
\right. \label{F-cot-n} \\
\widehat{\Psi}_n &=&
\left\{
\begin{array}{ll}
2 n  \mbox{e}^{ n b-i n a}, & n \ge 1, \\
                  0, & n \le 0.
\end{array}
\right. 
\end{eqnarray}

\section{Measurements and Observability}
\label{sec:observe}

We assume that the measurements have the form of the velocity recorded
continuously in time at some point $z_m = x_m + iy_m$ located away
from the vortex sheet, where without loss of generality we can assume
that $y_m > 0$, i.e.,
\begin{equation}
U\left(z_m\right) = \frac{1}{4\pi i}\int_0^{2\pi} \cot\left( \frac{z_m-z(\gamma,t)}{2}\right) \,d\gamma 
+ \sum_{k=1}^{2N_c} \frac{G_k}{4 \pi i} \cot\left(\frac{z_m - w_k}{2} \right).
\label{eq:U}
\end{equation}
This allows us to define the observation operator $\h$ as
\begin{equation}
\h[z]= 
\left[ 
\begin{aligned} 
  & \Re(U(z_m)) \\ 
- & \Im(U(z_m))
\end{aligned}
\right],
\label{eq:h}
\end{equation}
where the notation $\h[z]$ implies the dependence of the measurements
on the state variable $z$ which represents the instantaneous position
of the sheet. We will be primarily interested in the
state-to-measurements map represented by the first term on the RHS in
\eqref{eq:U}.

In order to assess the observability of the Birkhoff-Rott equation
\eqref{eq:BRc2} with measurements defined in \eqref{eq:h} linearized
around the flat-sheet configuration, we will follow {a}
formalism analogous to the approach employed in \S
\ref{sec:control_pv} and \S \ref{sec:control_ss} to study
controllability. {Observability implies that complete
  information about the state of the system can in the limit $t
  \rightarrow \infty$ be deduced from the available measurements.}
First, we need to obtain the linearization of the observation operator
\eqref{eq:h} and using ansatz \eqref{eq:zpert} we can write
\begin{equation}
\h[z]= \h[\tilde{z}] + \varepsilon \, \h'[\tilde{z}]\zeta + \mathcal{O}(\varepsilon^2),
\label{eq:dh}
\end{equation}
where
\begin{equation}
\h'[\tilde{z}]\zeta = 
\left[ 
\begin{aligned} 
  & \Re(U'(z_m)\zeta) \\ 
- & \Im(U'(z_m)\zeta)
\end{aligned}
\right]
\label{eq:h'}
\end{equation}
with
\begin{equation}
U'(z_m)\zeta = - \frac{1}{4\pi i} 
\int_0^{2\pi} \frac{\zeta(\gamma)}{2\sin^2\left(\frac{\gamma-z_k}{2}\right)}\, d\gamma.
\label{eq:U'}
\end{equation}
Assuming that the perturbation $\zeta$ is given by the Fourier series
\eqref{eq:zeta} truncated at the wavenumber $N$ and converted to the
real-valued representation as in \eqref{eq:dabdt}, expression
\eqref{eq:h'} can be approximated as
\begin{equation}
\h'[\tilde{z}]\zeta \approx \bH \X,
\label{eq:h'2}
\end{equation}
where $\bH = \left[ \bH_1 \ \bH_2 \ \ldots \ \bH_N \right]$ is the
observation matrix. System \eqref{eq:ABG_N} with measurements
\eqref{eq:h'2} is {then} declared observable if the following
condition holds \citep{s94}
\begin{equation}
\mbox{Rank}\left[ 
\begin{aligned}
& \bH \\ & \bH \A \\ & \cdots \\  & \bH \A^{N-1}
\end{aligned} \right] 
= 4N,
\label{eq:RH}
\end{equation}
which, given the block-diagonal structure of the matrix $\A$, cf.~\S
\ref{sec:KH}, is satisfied if {and only if} the rank condition
given below simultaneously holds for all individual blocks, i.e.,
\begin{equation}
\R_n^{H} = 
\left[
\begin{aligned}
& \bH_n \\ & \bH_n \A_n \\ & \bH_n \A_n^2 \\  & \bH_n \A_n^3
\end{aligned} 
\right] = 4, \quad n \ge 1.
\label{eq:RHn}
\end{equation}
Using expansions \eqref{eq:zeta} and \eqref{eq:f2} in \eqref{eq:U'},
we obtain
\begin{align}
U'(z_m)\zeta & = -\frac{i}{2} \sum_{n=-\infty}^\infty \widehat{\Psi}_{-n} \, \hzeta_n 
\nonumber \\
& = \sum_{n=1}^\infty n e^{-n y_m} \left[ 
\left( \sin(n x_m) - i \cos(n x_m) \right) \alpha_n +
\left( \cos(n x_m) + i \sin(n x_m) \right) \beta_n \right],
\label{eq:U'2}
\end{align}
where in the second equality we also used identity \eqref{F-sin2-p}.
After truncating the series at $n=N$ and rearranging it into a form
consistent with \eqref{eq:h'}, we obtain the following expression for
the $n$th block $\bH_n$ of the observation matrix
\begin{equation}
\bH_n = n e^{-n y_m} 
\begin{bmatrix}
& 0 & \cos(n x_m) & 0 & \sin(n x_m) \\
& 0 & -\sin(n x_m) & 0 & \cos(n x_m) 
\end{bmatrix}
\label{eq:H}.
\end{equation}
It can be verified using symbolic-algebra tools such as {\tt Maple}
that condition \eqref{eq:RHn} does indeed hold for $n \ge 1$, implying
the observability of the matrix pairs $\{ \A_n, \bH_n \}$, $n \ge 1$,
and hence the observability of system \eqref{eq:ABG_N} with the
measurements \eqref{eq:h'2} in the limit $N \rightarrow \infty$. We
add that, as is evident from the form of expression \eqref{eq:H},
modes with increasing wavenumbers $n$, while in principle observable,
leave only an exponentially decreasing footprint on the measurements.
Moreover, observability is lost when only one velocity component is
observed, cf.~\eqref{eq:h}.


\end{document}